\documentclass[11pt]{article}

\usepackage{a4wide}
\usepackage{multirow}
\usepackage{endfloat}
\usepackage{graphicx}
\usepackage{amsmath}
\usepackage{amssymb}
\usepackage{amsfonts}
\usepackage{natbib}
\usepackage{pstricks}
\usepackage{setspace} 
\usepackage{aeguill} 

\def\ee{I\mskip-7muE} 

\def\bx{\mbox{\boldmath $x$}}  

\def\bu{\mbox{\boldmath $u$}}

\def\bp{\mbox{\boldmath $p$}}
\def\bk{\mbox{\boldmath $k$}}

\def\btheta{\mbox{\boldmath $\theta$}}
\def\bbeta{\mbox{\boldmath $\beta$}}
\def\bSigma{\mbox{\boldmath $\Sigma$}}

\begin{document}

\title{CALCULATIONS OF SOBOL INDICES FOR THE GAUSSIAN PROCESS METAMODEL}

\author{Amandine MARREL$^{\ast,1}$, Bertrand IOOSS$^\dag$, B\'eatrice LAURENT$^\diamond$, Olivier ROUSTANT$^\ddag$}
\date{}

\maketitle

\begin{center}
  Submitted to: {\it Reliability Engineering and System Safety}\\
for the special SAMO 2007 issue

\vspace{0.2cm}

$^\ast$ CEA Cadarache, DEN/DTN/SMTM/LMTE, 13108 Saint Paul lez
  Durance, Cedex, France

$^\dag$ CEA Cadarache, DEN/DER/SESI/LCFR, 13108 Saint Paul lez
  Durance, Cedex, France

$^\diamond$ Institut de Mathématiques, Université de Toulouse (UMR 5219), France

$^\ddag$ Ecole des Mines de Saint-Etienne, France


\end{center}

\doublespacing

\abstract{ 
Global sensitivity analysis of complex numerical models can be performed by calculating variance-based importance measures of the input variables, such as the Sobol indices.
However, these techniques, requiring a large number of model evaluations, are often unacceptable for time expensive computer codes.
A well known and widely used decision consists in replacing the computer code by a metamodel,  predicting the model responses with a negligible computation time and rending straightforward the estimation of Sobol indices.
In this paper, we discuss about the Gaussian process model which gives analytical expressions of Sobol indices. 
Two approaches are studied to compute the Sobol indices: the first based on the predictor of the Gaussian process model and the second based on the global stochastic process model. 
Comparisons between the two estimates, made on analytical examples, show the superiority of the second approach in terms of convergence and robustness.
Moreover, the second approach allows to integrate the modeling error of the Gaussian process model by directly giving some confidence intervals on the Sobol indices.
These techniques are finally applied to a real case of hydrogeological modeling.}

\vspace{0.2cm}


\footnotetext[1]{Corresponding author: A. Marrel, Email: amandine.marrel@cea.fr\\
Phone: +33 (0)4 42 25 26 52, Fax: +33 (0)4 42 25 62 72
}

\clearpage

\section{INTRODUCTION}

Environmental risk assessment is often based on complex computer codes, simulating for instance an atmospheric or hydrogeological pollution transport.
These computer models calculate several output values (scalars or functions) which can depend on a high number of input parameters and physical variables.
To provide guidance to a better understanding of this kind of modeling and in order to reduce the response uncertainties most effectively, sensitivity measures of the input importance on the response variability can be useful (Saltelli et al. \cite{salcha00}, Kleijnen \cite{kle97}, Helton et al. \cite{heljoh06}).
However, the estimation of these measures (based on Monte-Carlo methods for example) requires a large number of model evaluations, which is unacceptable for time expensive computer codes.
This kind of problem is of course not limited to environmental modeling and can be applied to any simulation system.

To avoid the problem of huge calculation time in sensitivity analysis, it can be useful to replace the complex computer code by a mathematical approximation, called a response surface or a surrogate model or also a metamodel.
The response surface method (Box \& Draper \cite{boxdra87}) consists in constructing a function from few experiments, that simulates the behavior of the real phenomenon in the domain of influential parameters.
These methods have been generalized to develop surrogates for costly computer codes (Sacks et al. \cite{sacwel89}, Kleijnen \& Sargent \cite{klesar00}).
Several metamodels are classically used: polynomials, splines, generalized linear models, or learning statistical models like neural networks, regression trees, support vector machines (Chen et al. \cite{chetsu06}, Fang et al. \cite{fanli06}).

Our attention is focused on the Gaussian process model which can be viewed as an extension of the kriging principles (Matheron \cite{mat70},  Cressie \cite{cre93}, Sacks et al. \cite{sacwel89}). This metamodel which is characterized by its mean and covariance functions, presents several advantages: it is an exact interpolator and it is interpretable (not a black-box function). Moreover, numerous authors (for example, Currin et al. \cite{curmit91}, Santner et al. \cite{sanwil03}, Vazquez et al. \cite{vazwal05}, Rasmussen \& Williams \cite{raswil06}) have shown how this model can provide a statistical basis for computing an efficient predictor of code response. In addition to its efficiency, this model gives an analytical formula which is very useful for sensitivity analysis, especially for the variance-based importance measures, the so-called Sobol indices (Sobol \cite{sob93}, Saltelli et al. \cite{salcha00}).  To derive analytical expression of Sobol indices, Chen et al. \cite{chejin05} used tensor-product formulation  and Oakley \& O'Hagan \cite{oakoha04} considered the Bayesian formalism of Gaussian processes. 

We propose to compare these two analytical formulations of Sobol indices for the Gaussian process model: the first is obtained considering only the predictor, i.e. the mean of the Gaussian process model (Chen et al. \cite{chejin05}), while the second is obtained using all the global stochastic model (Oakley \& O'Hagan \cite{oakoha04}). In the last case, the estimate of a Sobol index is itself a random variable. Its standard deviation is available and we propose an original algorithm to  estimate its distribution.
Consequently, our method leads to build confidence intervals for the Sobol indices.
To our knowledge, this information has not been proposed before and can be obtained thanks to the analytical formulation of the Gaussian process model error.
This is particularly interesting in practice, when the predictive quality of the metamodel is not high (because of small learning sample size for example), and our confidence on Sobol index estimates via the metamodel is poor.

The next section briefly explains the Gaussian process modeling and the Sobol indices defined in the two approaches (predictor-only and global model). In section 3, the numerical computation of a Sobol index is presented. In the case of the global stochastic model, a procedure is developed to simulate its distribution. Section 4 is devoted to applications on analytical functions. First, comparisons are made between the Sobol indices based on the predictor and those based on the global model.
The pertinence of simulating all the distribution of Sobol indices is therefore evaluated. Finally, Sobol indices and their uncertainty are computed for a real data set coming from a hydrogeological transport model based on waterflow and diffusion dispersion equations. The last section provides some possible extensions and concluding remarks.


\section{SOBOL INDICES WITH GAUSSIAN PROCESS MODEL}\label{secgpmodel}

\subsection{Gaussian process model}

Let us consider $n$ realizations of a computer code. Each realization $y(\bx)$ of the computer code output corresponds to a d-dimensional input vector $\bx = (x_1,...,x_d)$. The $n$ input points corresponding to the code runs are called an experimental design and are denoted as $X_s = ( \bx^{(1)},...,\bx^{(n)} )$. The outputs will be denoted as $Y_s = (y^{(1)},...,y^{(n)})$ with $y^{(i)} = y(\bx^{(i)}), i=1,...,n$.
Gaussian process (Gp) modeling treats the deterministic response $y(\bx)$ as a realization of a random function $Y(\bx)$, including a regression part and a centered stochastic process. The sample space $\Omega$ denotes the space of all  possible outcomes $\omega$, which is usually  the Lebesgue-measurable set of real numbers. The Gp is defined on $ R^d \times \Omega $ and can be written as:
\begin{equation}\label{eqfirst}
 Y ( \bx, \omega ) = f ( \bx ) + Z ( \bx, \omega) .
\end{equation}
In the following, we use indifferently the terms Gp model and Gp metamodel.

The deterministic function $f(\bx)$ provides the mean approximation of the computer code. Our study is limited to the parametric case where the function $f$ is a linear combination of elementary functions.
Under this assumption, $f(\bx)$ can be written as follows:
\[ f ( \bx ) = \sum_{j = 0}^k \beta_j f_j ( \bx ) = F ( \bx ) \bbeta , \]
where $\bbeta = [ \beta_0, \ldots, \beta_k ]\text{}^t $ is the regression parameter vector, $f_j$ ($j=1, \ldots, k$) are basis functions and
 $F ( \bx ) =  [ f_0 ( \bx ), \ldots, f_k ( \bx ) ]$ is the corresponding regression matrix. 
In the case of the one-degree polynomial regression, $(d+1)$ basis functions are used:
 \[
    \left\{ \begin{array}{l}
      f_0 ( \bx ) = 1 ,  \\
      f_i ( \bx ) = x_i \;\text{ for }\; i=1, \ldots,d .
    \end{array}  \right.
  \]

In our applications, we use this one-degree polynomial as the regression part in order to simplify all the analytical numerical computation of sensitivity indices. This can be extended to other bases of regression functions.  Without prior information on the relationship between the output and the inputs, a basis of one-dimensional functions is recommended to simplify the computations in sensitivity analysis and to keep one of the most advantages of Gp model (Martin \& Simpson \cite{marsim05}). 

The stochastic part $Z(\bx, \omega)$ is a Gaussian centered process fully characterized by its covariance function:
$\mbox{Cov}_{\Omega} ( Z ( \bx,\omega ), Z ( \bu,\omega ) ) = \sigma^2 R ( \bx, \bu ),$
where $\sigma^2$ denotes the variance of $Z$ and $R$ is the correlation function that provides interpolation and spatial correlation properties.
To simplify, a stationary process $(Z(\bx, \omega))$ is considered, which means that the correlation between $Z(\bx,\omega)$ and $Z(\bu,\omega)$ is a function of the difference between $\bx$ and $\bu$. Moreover, our study is restricted to a family of correlation functions that can be written as a product of one-dimensional correlation functions:
\begin{equation}\label{cov_prod}
 \mbox{Cov}_{\Omega} ( Z ( \bx,\omega ), Z ( \bu,\omega ) )  = \sigma^2 R ( \bx - \bu ) = \sigma^2 \prod_{l = 1}^d R_l ( x_l - u_l ) . 
\end{equation}
This form of correlation function is particularly well adapted to get some simplifications of the integrals in the future analytical developments: in the case of independent inputs, it implies the computation of only one or two-dimensional integrals to compute the Sobol indices. Indeed, as described in section \ref{sec_sobol_comput}, the application and the computation of the Sobol index formulae are simplified when the correlation function has the form of a  one-dimensional product (Santner et al. \cite{sanwil03}).

Among other authors, Chil\`es \& Delfiner \cite{chidel99} and Rasmussen \& Williams \cite{raswil06} give a list of correlation functions with their advantages and drawbacks.
Among all these functions, our attention is devoted to the generalized exponential correlation function:
\[ R_{\btheta, \bp} ( \bx - \bu ) = \prod_{l = 1}^d \exp ( - \theta_l |x_l - u_l |^{p_l} ) \text{ with } \theta_l \geq 0  \text{ and } 0 < p_l \leq 2 ,\]
where  $\btheta = [ \theta_1, \ldots, \theta_d ]\text{}^t $ and $\bp =  [ p_1, \ldots, p_d ]\text{}^t $ are the correlation parameters.
This choice is motivated by the derivation and regularity properties of this function.
Moreover, within the range  of covariance parameters values,  a wide spectrum of  shapes are possible: for example $p = 1$ gives the exponential correlation function while $p=2$ gives the Gaussian correlation function.

\subsection{Joint and conditional distributions} \label{secjoint}

Under the hypothesis of a Gp model, the learning sample $Y_s$ follows a multivariate normal distribution $p_{\Omega}(Y_s \left|X_s \right )$:
\[ p_{\Omega}( Y_s,\omega  | X_s  ) = \mathcal{N} \left(F_s \bbeta, \bSigma_s \right) ,\]
where $F_s = [ F ( \bx^{(1)} )\textrm{}^t , \ldots, F ( \bx^{(n)}\textrm{}^t )] $ is the regression matrix and 
\[ \bSigma_s = \sigma^2 R_{\btheta, \bp} \left( \bx^{(i)} - \bx^{(j)} \right)_{i, j = 1 \ldots n} \]
is the covariance matrix.

If a new point $\bx^{\ast} = (x^{\ast}_1,...,x^{\ast}_d)$ is considered, the joint probability distribution of $(Y_s, Y(\bx^{\ast},\omega ))$ is:
\begin{equation}
  p_{\Omega}( Y_s, Y(\bx^{*},\omega ) | X_s, \bx^{\ast},\bbeta,\sigma,\btheta, \bp ) = \mathcal{N} \left( \left[ \begin{array}{c}F_s \\ F(\bx^{*}) \end{array}\right] \bbeta,    \left[ \begin{array}{c c} \bSigma_s &  \bk(\bx^{*}) \\ \bk(\bx^{*})\text{}^t & \sigma^2 \end{array}\right]  \right) ,
\end{equation}
with
  \begin{equation} 
\begin{array}{lll}
\bk(\bx^{*}) & = &( \: \mbox{Cov}_{\Omega} (y^{(1)},Y(\bx^{*},\omega)), \ldots, \mbox{Cov}_{\Omega} (y^{(n)},Y(\bx^{*},\omega)) \: ) \text{}^t  \\
& = & \sigma^2 (  \: R_{\btheta, \bp} (\bx^{(1)},\bx^{*}), \ldots, R_{\btheta, \bp} (\bx^{(n)},\bx^{*})  \: )  \text{}^t .   
 \end{array} 
\end{equation}  

By conditioning this joint distribution on the learning sample, we can readily obtain the conditional distribution of $Y(\bx^{\ast},\omega)$ which is Gaussian (von Mises \cite{von64}):
 \begin{equation}
 \begin{array}{l}
  p_{\Omega}(Y(\bx^{*},\omega) | Y_s,X_s, \bx^{\ast},\bbeta,\sigma,\btheta, \bp) \\
  = \mathcal{N} \left( \ee_{\Omega} [  Y(\bx^{*},\omega) | Y_s,X_s, \bx^{\ast},\bbeta,\sigma,\btheta, \bp] , \mbox{Var}_{\Omega}[  Y(\bx^{*},\omega) | Y_s,X_s, \bx^{\ast} ,\bbeta,\sigma,\btheta, \bp] \right) ,
\end{array}
\end{equation}
with
 \begin{eqnarray}
 \displaystyle \ee_{\Omega} [  Y(\bx^{*},\omega) | Y_s,X_s, \bx^{\ast},\bbeta,\sigma,\btheta, \bp ] & = &  F(\bx^{*})\bbeta  +  \bk(\bx^{*})  \text{}^t \bSigma_s^{-1} (Y_s - F_s\bbeta) , \label{eq_esperance}  \\ 
\displaystyle \mbox{Var}_{\Omega}[ Y(\bx^{*},\omega) | Y_s,X_s, \bx^{\ast},\bbeta,\sigma,\btheta, \bp] & = & \sigma^2   -  \bk(\bx^{*}) \text{}^t  \bSigma_s^{-1} \bk(\bx^{*}) .  \label{eq_variance}
   \end{eqnarray}

The conditional mean of Eq. (\ref{eq_esperance}) is  used as a predictor. The conditional variance formula of Eq. (\ref{eq_variance}) corresponds to the mean squared error (MSE) of this predictor and is also known as the kriging variance. As we obtained the distribution for a new point conditionally to the learning sample, we can consider the covariance between two new sites. A Gp conditional to the learning sample is obtained and denoted as follows:
\begin{equation}\label{eqPG}
\begin{array}{rl}
\displaystyle ( Y  | Y_s,X_s,\bbeta,\sigma,\btheta, \bp ) \sim \mbox{Gp} ( & \displaystyle \ee_{\Omega} [  Y(\bx^{*},\omega ) | Y_s,X_s, \bbeta,\sigma,\btheta, \bp ] ,\\
& \displaystyle \mbox{Cov}_{\Omega} (Y(\bx^{*},\omega ) , Y(\bu^{*},\omega ) | Y_s,X_s, \bx^{\ast},\bbeta,\sigma,\btheta, \bp  ) )
\end{array} \end{equation}
with the same expression for the conditional mean than Eq. (\ref{eq_esperance}) and
 \begin{equation}\label{eq_PG_cond_cov} 
\displaystyle \mbox{Cov}_{\Omega}\left(Y(\bx^{*},\omega ) , Y(\bu^{*},\omega ) | Y_s,X_s,\bbeta,\sigma,\btheta, \bp  \right) = \sigma^2 \left( R_{\btheta, \bp} (\bx^{*},\bu^{*}) -\bk(\bx^{*}) \text{}^t  \bSigma_s^{-1} \bk(\bu^{*})  \right).  
   \end{equation}

The conditional Gp model (\ref{eqPG}) provides an analytical formula which can be directly used for sensitivity analysis, and more precisely to compute the Sobol indices. To simplify the notations, the conditional Gp $\left( Y | Y_s,X_s,\bbeta,\sigma,\btheta, \bp \right)$ will now be written in a simplified form: $  Y_{\mbox{Gp}| Y_s,X_s} (X,\omega)$.
 
\subsection{Sobol indices}\label{secSoboldef}

Methods based on variance decomposition aim at determining the part of the variance of the output $Y(\bx)$ resulting from each variable $\bx_i, i=1,\ldots,d$. 
A global measure of the sensitivity of $Y(\bx)$ to each input $\bx_i$ is given by the first order Sobol index (Sobol \cite{sob93}, Saltelli et al. \cite{salcha00}):
\[ S_i = \frac{\mbox{Var}_{X_i} [\ee_{X_1,\ldots, X_d } Y | X_i]}{\mbox{Var}_{X_1,\ldots, X_d} [Y]} \text{ for } i=1,\ldots,d. \] 

These indices have been defined for deterministic functions $Y$ of the inputs $X_1,\ldots, X_d$ but, in the case of the conditional Gp model, we have a stochastic function of the inputs. 
A first solution is applying the Sobol index formula to the predictor, i.e. the mean of the conditional Gp (Eq. (\ref{eq_esperance})) which is a deterministic function of the inputs. Analytical calculations are developed by Chen et al. \cite{chejin05}. The second approach that we consider consists in using the whole global conditional Gp by taking into account not only the mean of conditional Gp model but also its covariance structure as Oakley \& O'Hagan \cite{oakoha04} did. In this case, when the Sobol definition is applied to the global Gp model, a random variable is obtained and constitutes a new sensitivity measure. Its expectation can be then considered as a sensitivity index. Its variance and more generally its distribution can then be used as an indicator of sensitivity index accuracy. 

To sum up, the two approaches can be defined as follows:
\begin{itemize}
\item \underline{Approach 1:} Sobol indices computed with the predictor-only
\begin{equation}\label{eq_sobol_approach1}  
 S_i =  \frac{\mbox{Var}_{X_i} \ee_{X_1,\ldots, X_d }  [ \ee_{\Omega} [Y_{\mbox{Gp}| Y_s,X_s} (X,\omega) ] | X_i]}{\mbox{Var}_{X_1,\ldots, X_d} \ee_{\Omega}[Y_{\mbox{Gp}| Y_s,X_s} (X,\omega)  ]}  \text{ for } i=1,\ldots,d.
\end{equation}
\item \underline{Approach 2:} Sobol indices computed with the global Gp model
\begin{equation}\label{eq_sobol_approach2}  
\tilde{S_i}(\omega) =  \frac{\mbox{Var}_{X_i} \ee_{X_1,\ldots, X_d } [Y_{\mbox{Gp}| Y_s,X_s} (X,\omega) | X_i]}{\ee_{\Omega} \mbox{Var}_{X_1,\ldots, X_d} [Y_{\mbox{Gp}| Y_s,X_s}  (X,\omega) ]}  \text{ for } i=1,\ldots,d. 
\end{equation}
$ \tilde{S_i}(\omega)$ is then a random variable; its mean can be considered as a sensitivity index and its variance as an indicator of its accuracy:
\begin{equation}\label{eq_sobol_approach2b}  
 \left\{ \begin{array}{l}
 \displaystyle \mu_{\tilde{S_i}}  = \frac{\ee_{\Omega} \mbox{Var}_{X_i} \ee_{X_1,\ldots, X_d } [Y_{\mbox{Gp}| Y_s,X_s} (X,\omega) | X_i]}{\ee_{\Omega} \mbox{Var}_{X_1,\ldots, X_d} [Y_{\mbox{Gp}| Y_s,X_s} (X,\omega)  ]}  \text{ for } i=1,\ldots,d.  \\
\\
 \displaystyle \sigma^2_{\tilde{S_i}}  = \frac{\mbox{Var}_{\Omega} \mbox{Var}_{X_i} \ee_{X_1,\ldots, X_d } [Y_{\mbox{Gp}| Y_s,X_s} (X,\omega) | X_i]}{ ( \ee_{\Omega} \mbox{Var}_{X_1,\ldots, X_d} [Y_{\mbox{Gp}| Y_s,X_s} (X,\omega)  ] )^2 }  \text{ for } i=1,\ldots,d. 
   \end{array}  \right. 
\end{equation}
\end{itemize}

Our work focuses on the computation and the study of the sensitivity indices defined following the two approaches, respectively $ S_i$ and $\mu_{\tilde{S_i}}$. We will also propose a methodology to numerically simulate the probability distribution of $\tilde{S_i}$. Then, a study to compare the accuracy and the robustness of the two indices is made on several test functions and the use of the distribution of $ \tilde{S_i}$ is illustrated to build confidence intervals. 


\section{IMPLEMENTATION OF SOBOL INDICES} \label{sec_implement}

\subsection{Estimation of Gp parameters}

First of all, to build the conditional Gp defined by Eq. (\ref{eqPG}), regression and correlation parameters (often called hyperparameters) have to be determined. Indeed, the Gp model is characterized by the regression parameter vector $\bbeta$, the correlation parameters $(\btheta,\bp)$ and the variance parameter $\sigma^2$. The maximum likelihood method is commonly used to estimate these parameters from the learning sample $(X_s,Y_s)$. 

Several algorithms have been proposed in previous papers to numerically solve the maximization of likelihood.
Welch at al. \cite{welbuc92} use the simplex search method and introduce a kind of forward selection algorithm in which correlation parameters are added step by step to increase the log-likelihood function.
In Kennedy and O'Hagan's GEM-SA software (O'Hagan \cite{oha06}), which uses the Bayesian formalism, the posterior distribution of hyperparameters is maximized, using a conjugate gradient method (the Powel method is used as the numerical recipe).
The DACE Matlab free toolbox (Lophaven et al. \cite{lopnie02}) uses a powerful stochastic algorithm  based on the Hooke \& Jeeves method (Bazaraa et al. \cite{bazshe93}), which requires a starting point and some bounds to constrain the optimization.
In complex applications, Welch's algorithm reveals some limitations and for complex model with high dimensional input, GEM-SA and DACE software cannot be applied directly on data including all the input variables.
To solve this problem, we use a sequential version (inspired by Welch's algorithm) of the DACE algorithm.
It is based on the step by step inclusion of input variables (previously sorted).
This methodology, described in details in Marrel et al. \cite{marioo07}, allows progressive parameter estimation by input variables selection both in the regression part and in the covariance function.

\subsection{Computation of Sobol indices for the two approaches}\label{sec_sobol_comput}

To perform a variance-based sensitivity analysis for time consuming computer models,
some authors propose to approximate the computer code by a
metamodel (neural networks in Martin \& Simpson \cite{marmas03}, polynomials in Iooss et al. \cite{ioovan06}, boosting regression trees in Volkova et al. \cite{volioo07}).  For metamodels with sufficient prediction
capabilities, the bias due to the use of the metamodel instead of the
true model is negligible (Jacques \cite{jac05}).  
The metamodel's predictor have to be evaluated a large number of times to compute Sobol indices via Monte Carlo methods. 
Recent works based on polynomial chaos expansions (Sudret \cite{sud07}) have used the special form of this orthogonal functions expansion to derive analytical estimation of Sobol indices.
However, the modeling error of this metamodel is not available and then has not been integrated inside the Sobol index estimates.

The conditional Gp metamodel provides an analytic formula which can be easily used for sensitivity analysis in an analytical way. Moreover, in the case of independent inputs and with a covariance which is a product of one-dimensional covariances (Eq. (\ref{cov_prod})), the analytical formulae of  $S_i$ and $\mu_{\tilde{S_i}}$ (respectively Eqs. (\ref{eq_sobol_approach1}) and (\ref{eq_sobol_approach2b})) lead to numerical integrals, more precisely to respectively one-dimensional and two-dimensional integrals. The accuracy of these numerical integrations can be easily controlled and are less computer time expensive than Monte Carlo simulations. Few analytical developments of Sobol indices computation (for $S_i$, $\mu_{\tilde{S_i}}$ and  $\ \sigma^2_{\tilde{S_i}}$) can be found in Oakley \& O'Hagan \cite{oakoha04}.

\subsection{Simulation of the distribution of $\tilde{S_i}$}\label{sec_sim_distrib}

For the second approach  where $\tilde{S_i}$ is a random variable, the distribution of $\tilde{S_i}$ is not directly available. By taking the mean related to all the inputs except $X_i$, the main effect of $X_i$ is defined and denoted $A(X_i,\omega)$:
$$ A(X_i,\omega) = \ee_{X_1,\ldots, X_d } [Y_{\mbox{Gp}| Y_s,X_s} (X,\omega) | X_i].$$
 $A(X_i,\omega)$ is still a Gaussian process defined on $ R \times \Omega $ and characterized by its mean and covariance which can be determined in an analytical way by integrating the Gp model over all the inputs except $X_i$. In the case of independent inputs, one-dimensional integrals are obtained and can be numerically computed. Then, to obtain the Sobol indices, we consider the variance related to $X_i$ of the Gaussian process defined by the centered main effect. This variance is written
 $$ \int_{a_i}^{b_i} \left(  A(x_i,\omega) - \int A(x_i,\omega)d\eta_{x_i} \right)^2 d\eta_{x_i}$$
with $d\eta_{x_i}$ the probability density function of the input $X_i$ defined on $[a_i \, ; \, b_i]$.
This last expression is a one-dimensional random integral which has to be discretized and approximated by simulations. 

The discretization of this random integral over the space of $X_i$ leads to a Gaussian vector of $n_{\mbox{\tiny dis}}$ elements:
 $$ V_{n_{\mbox{\tiny dis}}} (\omega) = \left( A(a_i,\omega),\; A(a_i + \frac{(b_i-a_i)}{n_{\mbox{\tiny dis}}} ,\omega), \; \ldots, \; A(a_i + \frac{(n_{\mbox{\tiny dis}}-1)}{n_{\mbox{\tiny dis}}}(b_i-a_i) ,\omega),\;  A(b_i,\omega) \right) \text{}^t. $$
The mean and covariance matrix of this vector are computed using those of the Gaussian process $A(X_i,\omega)$. The random vector $ V_{n_{\mbox{\tiny dis}}}$ is then multiplied by the matrix related to the numerical scheme used to compute the integral (rectangle or trapeze method, Simpson's formula ...). The Gaussian vector obtained from this multiplication is denoted $\tilde{V}_{n_{\mbox{\tiny dis}}}$. 
To simulate it, we use the well known simulation method based on the Cholesky factorisation of the covariance matrix (Cressie \cite{cre93}).
We simulate a $n_{\mbox{\tiny dis}}$-size centered and reduced Gaussian vector and multiply it by the triangular matrix from the Cholesky decomposition. Then, an evaluation of the random integral which constitutes a realization of $\tilde{S_i}$ is computed from the simulation of the vector $\tilde{V}_{n_{\mbox{\tiny dis}}}$. This operation is done $k_{\mbox{\tiny sim}}$ times to obtain a probability distribution for $\tilde{S_i}$. It can be noted, that only one Cholesky factorization of the covariance matrix of the $n_{\mbox{\tiny dis}}$-size vector is necessary, and used for all the $k_{sim}$ simulations of $\tilde{S_i}$. To determine if the discretization number $n_{\mbox{\tiny dis}}$ and the number of simulations $k_{\mbox{\tiny sim}}$ are sufficient, the convergence of the mean and variance of $\tilde{S_i}$ can be studied. Indeed, their values can be easily computed following their analytical expressions (\ref{eq_sobol_approach2}).

\section{APPLICATIONS} \label{secappli}

\subsection{Comparison of $S_i$ and $\mu_{\tilde{S_i}}$}

To compare and study the behavior of the two sensitivity indices $S_i$ and $\mu_{\tilde{S_i}}$, we consider several test functions where the true values of Sobol indices are known. Comparisons between the two approaches are performed in terms of metamodel predictivity, i.e. relatively to the accuracy of the Gp model, constructed from a learning sample. This accuracy is represented by the predictivity coefficient $Q_2$. It corresponds to the classical coefficient of determination $R^2$ for a test sample, i.e. for prediction residuals:
 \[ Q_2 ( Y, \hat{Y} ) = 1 - \frac{\sum_{i = 1}^{n_{\mbox{\tiny test}}} \left( Y_i - \hat{Y}_i \right)^2}{\sum_{i = 1}^{n_{\mbox{\tiny test}}} \left( \bar{Y} - Y_i \right)^2}, \]
  where $Y$ denotes the $n_{\mbox{\tiny test}}$ true observations of the test set and $\bar{Y}$ is their empirical mean. $\hat{Y}$ represents the Gp model predicted values.
To obtain different values of $Q_2$, we simulate different learning samples with varying size $n$. For each size $n$, a Latin Hypercube Sample of the inputs is simulated (McKay et al. \cite{mckbec79}) to give the matrix $X_s$ ($n$ rows, $d$ columns).
Then, the test function is evaluated on the $n$ data points to constitute $(X_s,Y_s)$ and a conditional Gp model is built on each learning sample. For each Gp model built, the predictivity coefficient $Q_2$ is estimated on a new test sample of size 10000 and the two sensitivity indices $S_i$ and $\mu_{\tilde{S_i}}$ are computed. 
For each value of the learning sample size $n$, all this procedure, i.e. Gp modeling and estimation of sensitivity indices, is done $100$ times.
Consequently, the empirical mean, 0.05-quantile and 0.95-quantile of $S_i$ and $\mu_{\tilde{S_i}}$ are computed for same values of learning sample size $n$, and similar approximate values of $Q_2$.

\subsection{Test on the g-function of Sobol}\label{sec_comp_gSobol}

First, an analytical function called the g-function of Sobol is used to compare the Sobol indices $S_i$ based on the predictor and the Sobol indices $\mu_{\tilde{S_i}}$ based on the global Gp model. The g-function of Sobol is defined for $d$ inputs $(X_1,\ldots,X_d)$ uniformly distributed on $\left[ 0,1 \right]^{d}$:       
 \[
        g_{\mbox{\tiny Sobol}} (X_1, \ldots, X_d) = \prod _{k=1}^{d} g_k (X_k) \text{ where } g_k (X_k) = \frac{\left| 4X_k - 2 \right| + a_k}{1 + a_k}  \text{ and }
a_k \geq 0.
\]
Because of its complexity (considerable nonlinear and non-monotonic relationships) and to the availability of analytical sensitivity indices, it is a well known test example in the studies of global sensitivity analysis algorithms (Saltelli et al. \cite{salcha00}).
The importance of each input $X_k$ is represented by the coefficient $a_k$. The lower this coefficient $a_k$, the more significant the variable $X_k$. The theoretical values of first order Sobol indices are known:
\[ 
S_i = \frac{\frac{1}{3(1+a_i)^2}}{\prod _{k=1}^{d}\frac{1}{3(1+a_k)^2}}  \text{ for } i=1,\ldots,d. 
\]
For our analytical test, we choose $d = 5$ and $a_k = k  \text{ for } k=1,\ldots,5. $.

\begin{figure}[ht]
\begin{tabular}{ccc}
\includegraphics[height=5.cm,width=5.cm]{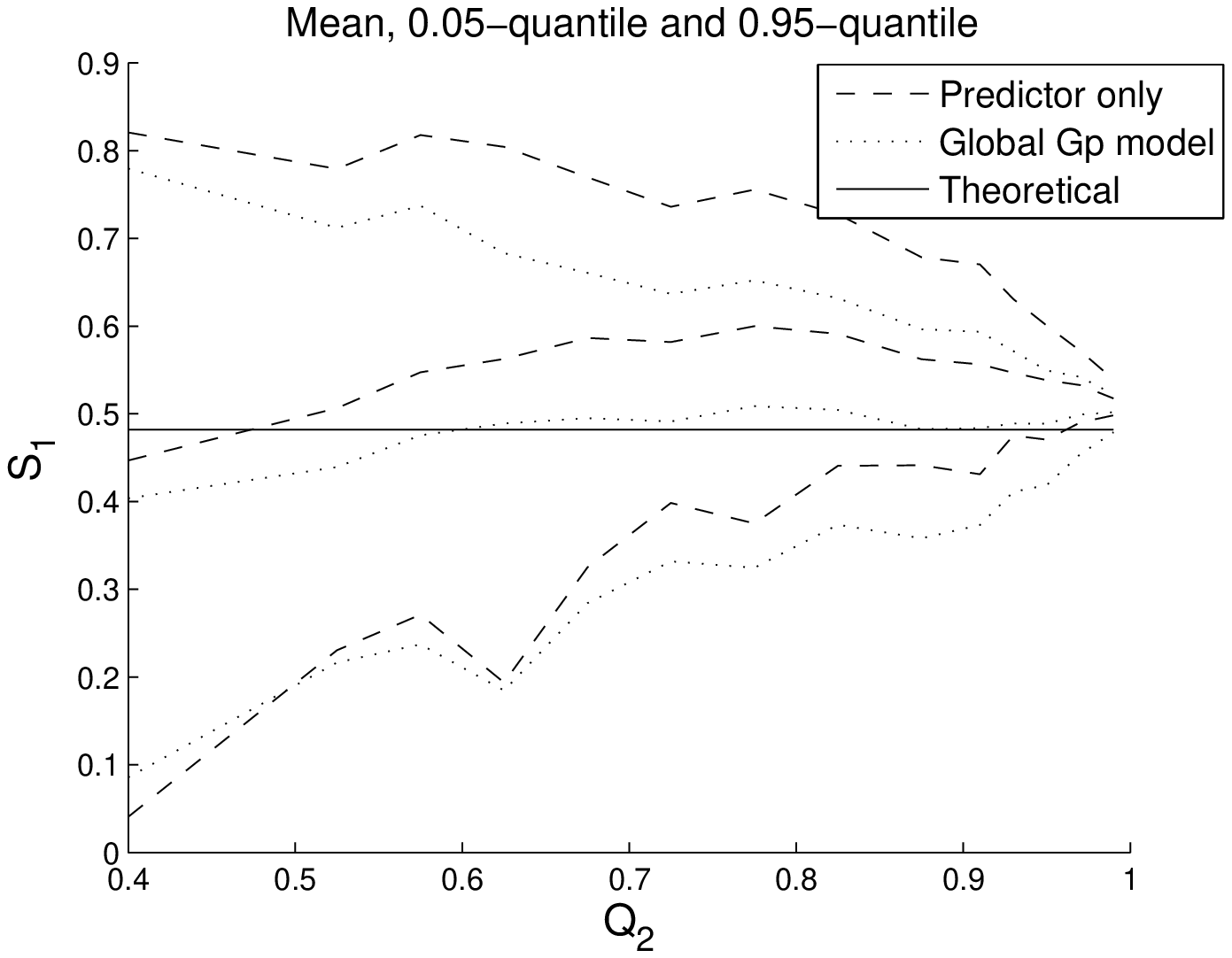} &
\includegraphics[height=5.cm,width=5.cm]{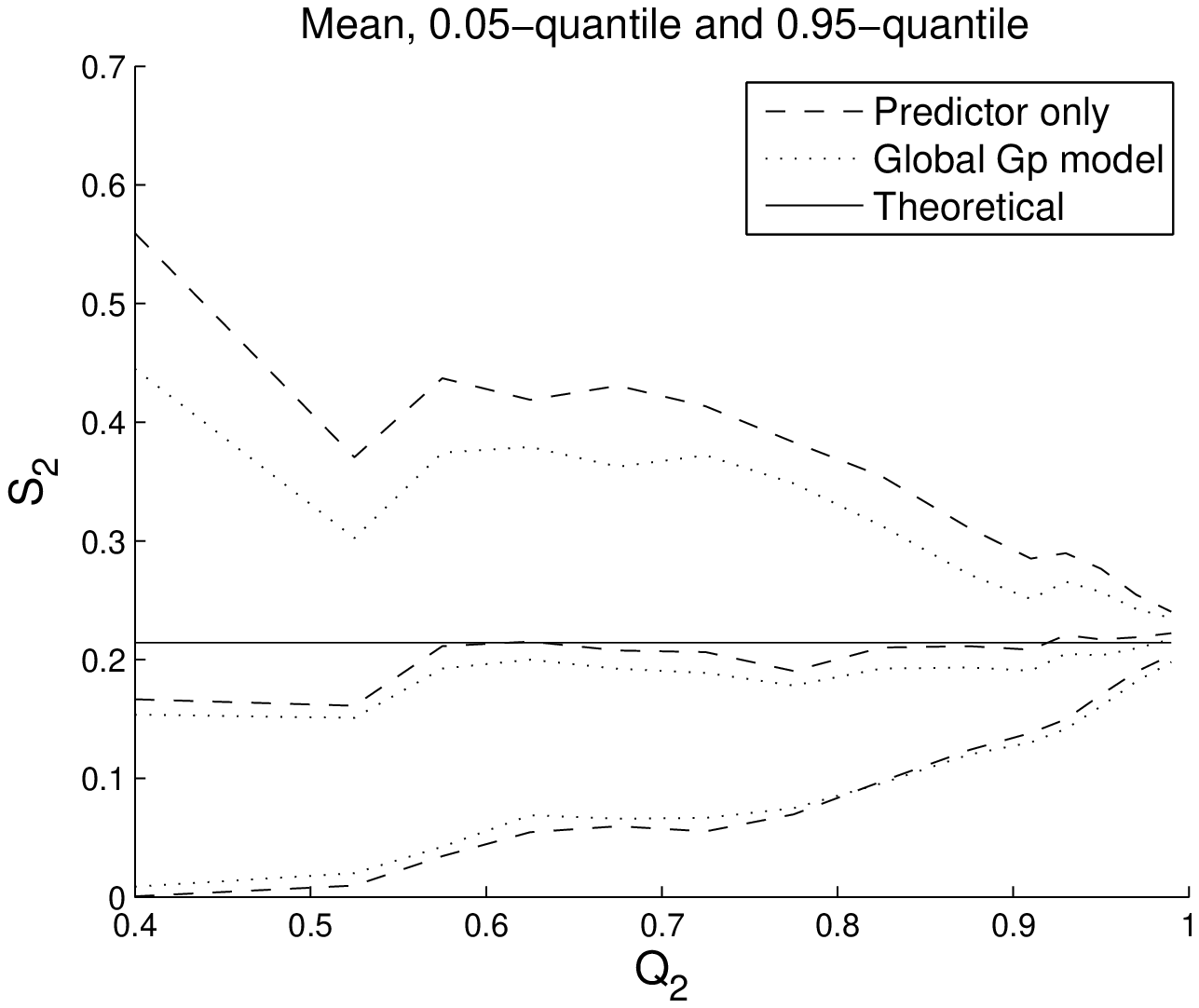} &
\includegraphics[height=5.cm,width=5.cm]{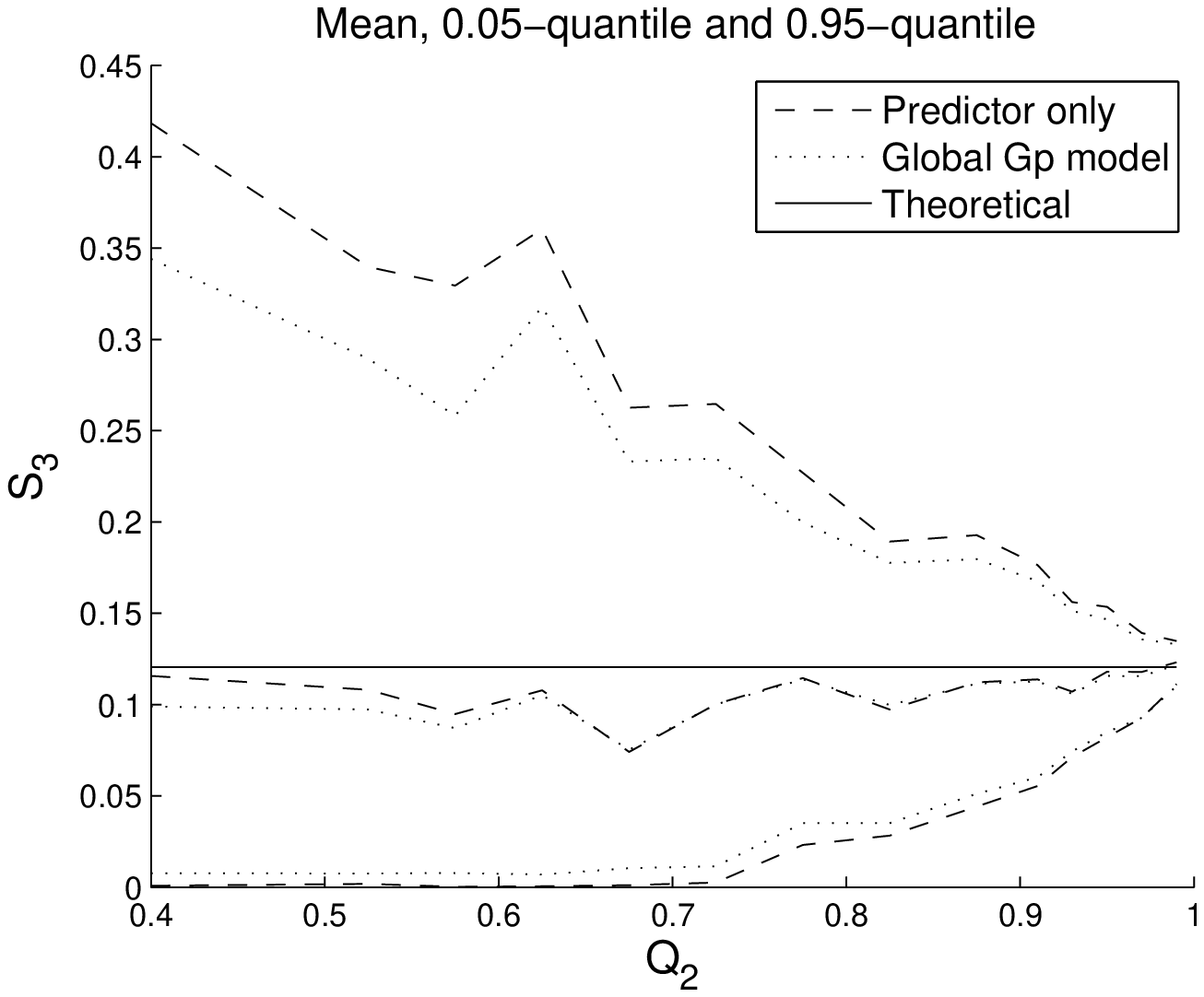} \\
\includegraphics[height=5.cm,width=5.cm]{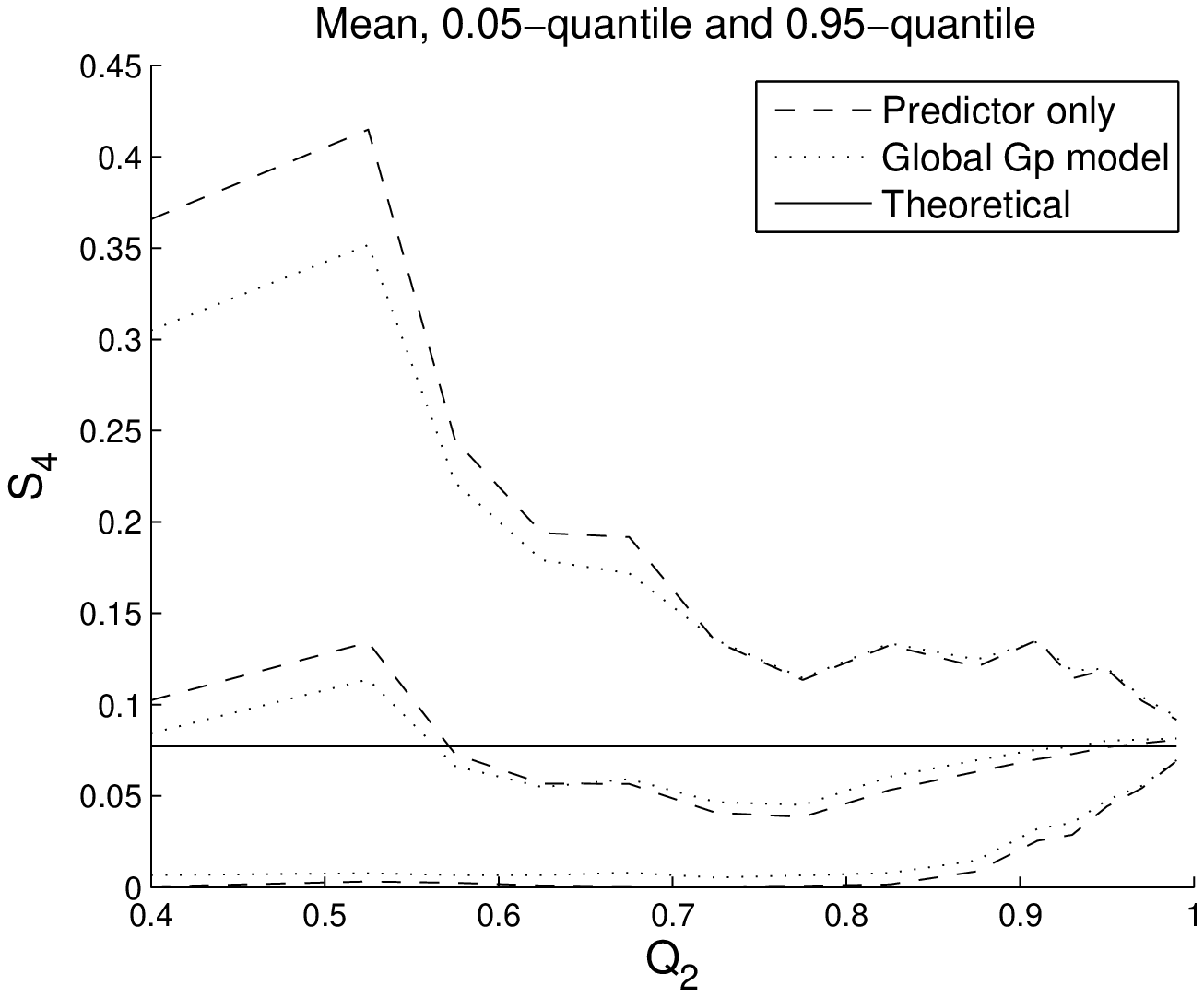} &
\includegraphics[height=5.cm,width=5.cm]{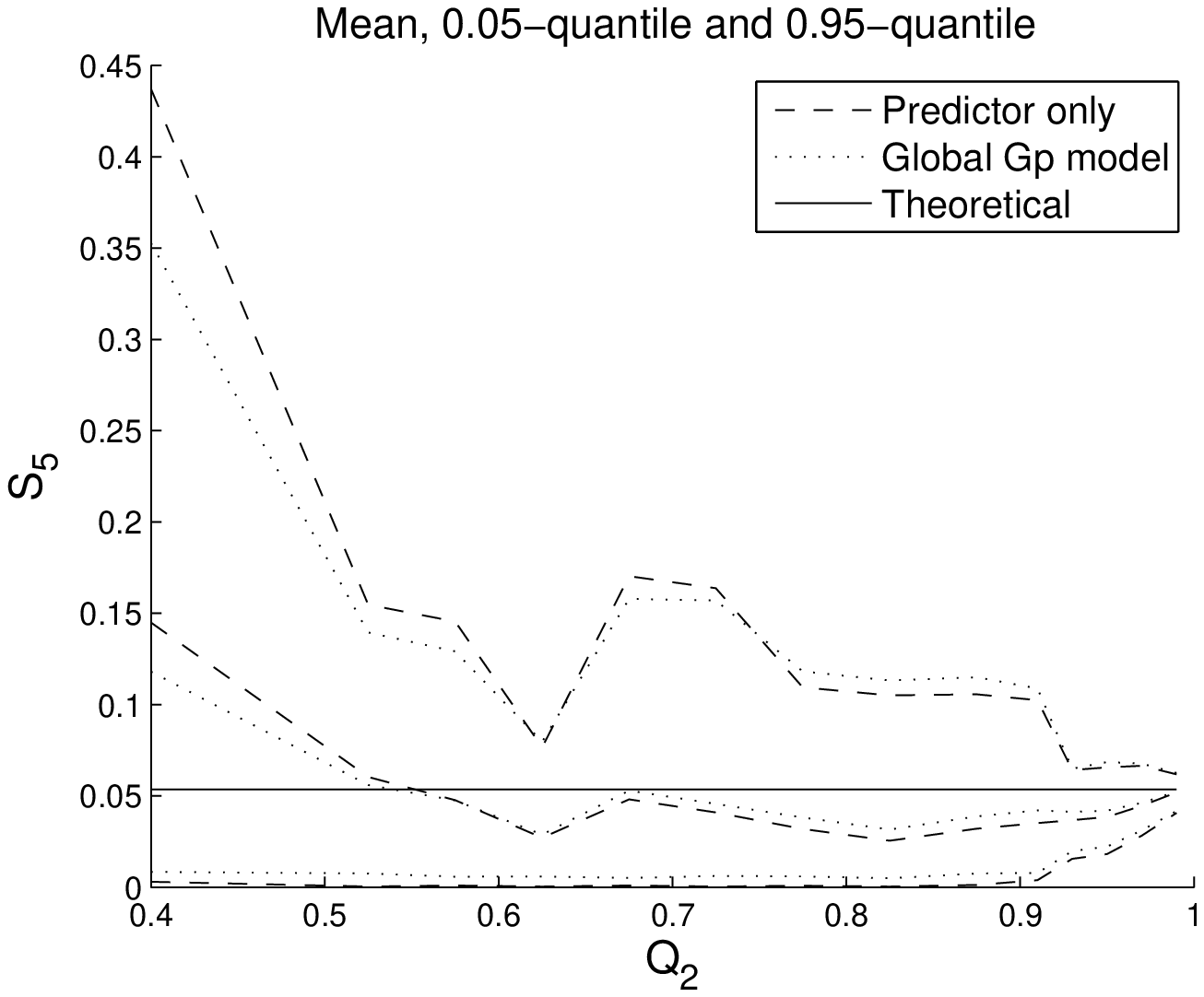} 
\end{tabular}
\caption{Convergence of sensitivity indices in function of the predictivity coefficient $Q_2$ (g-Sobol function).}
\label{fig_gSobol_Q2}
\end{figure}

Let us recall that we study only first order sensitivity indices.
For each input $X_i$, the convergence of $S_i$ and $\mu_{\tilde{S_i}}$ in function of the predictivity coefficient $Q_2$ is illustrated in figure \ref{fig_gSobol_Q2}. 
The convergence of sensitivity index estimates to their exact values in function of the metamodel predictivity is verified.
In practical situations, a metamodel with a predictivity lower than $0.7$ is often considered as a poor approximation of the computer code.
Table \ref{connec_n_Q2} shows the connection between the learning sample size $n$ and the predictivity coefficient $Q_2$. As the simulation of a learning sample and its Gp modeling are done $100$ times for each value of $n$, the mean and the standard deviation of $Q_2$ are indicated.
 \begin{table}[ht]
\begin{center}
\begin{tabular}{ccc}
\hline
Learning sample size $n$ & \multicolumn{2}{c}{Predictivity coefficient $Q_2$} \\
\cline{2-3}
   & Mean & Standard deviation \\
\hline
25 & 0.67  & 0.21\\
\hline
35  & 0.88  & 0.09 \\
\hline
45 & 0.96  & 0.02 \\
\hline
55 & 0.98  &  0.01\\
\hline
65 & 0.98  &  $6. 10^{-3}$ \\
\hline
75 & 0.99  &  $4. 10^{-3}$ \\
\hline
85 &   0.99 &  $3. 10^{-3}$ \\
\hline
95 &   0.99 &  $2. 10^{-3}$ \\
\hline
\end{tabular}
\end{center}
\caption{Connection between the learning sample size $n$ and the predictivity coefficient $Q_2$  (g-Sobol function).}
\label{connec_n_Q2}
\end{table}  
Figure \ref{fig_gSobol_Q2} also shows how the global Gp model outperforms the predictor-only model by showing smaller confidence intervals for the five sensitivity indices.
 
\begin{figure}[ht]
\begin{tabular}{cc}
\includegraphics[height=7.cm,width=7.cm]{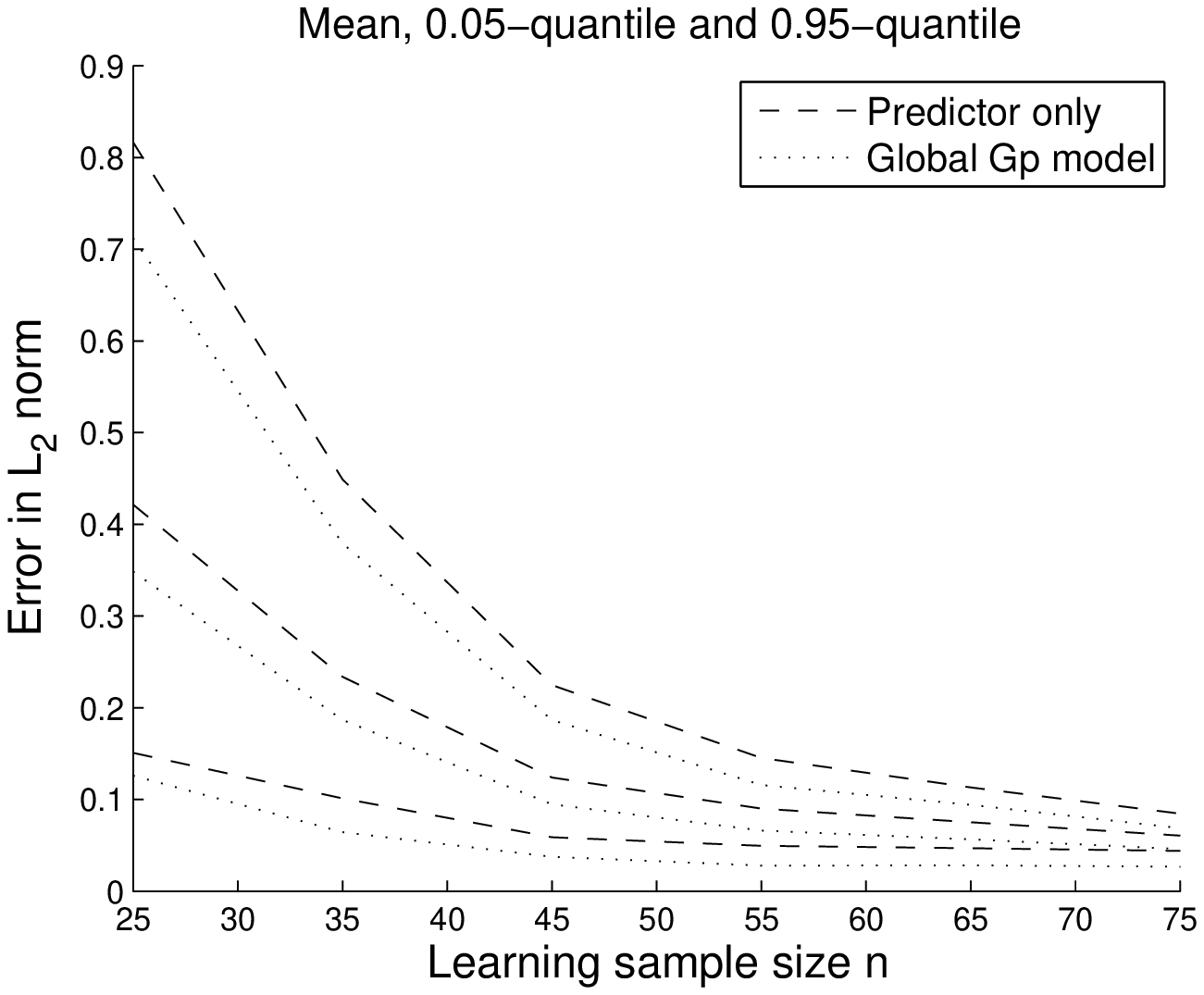} &
\includegraphics[height=7.cm,width=7.cm]{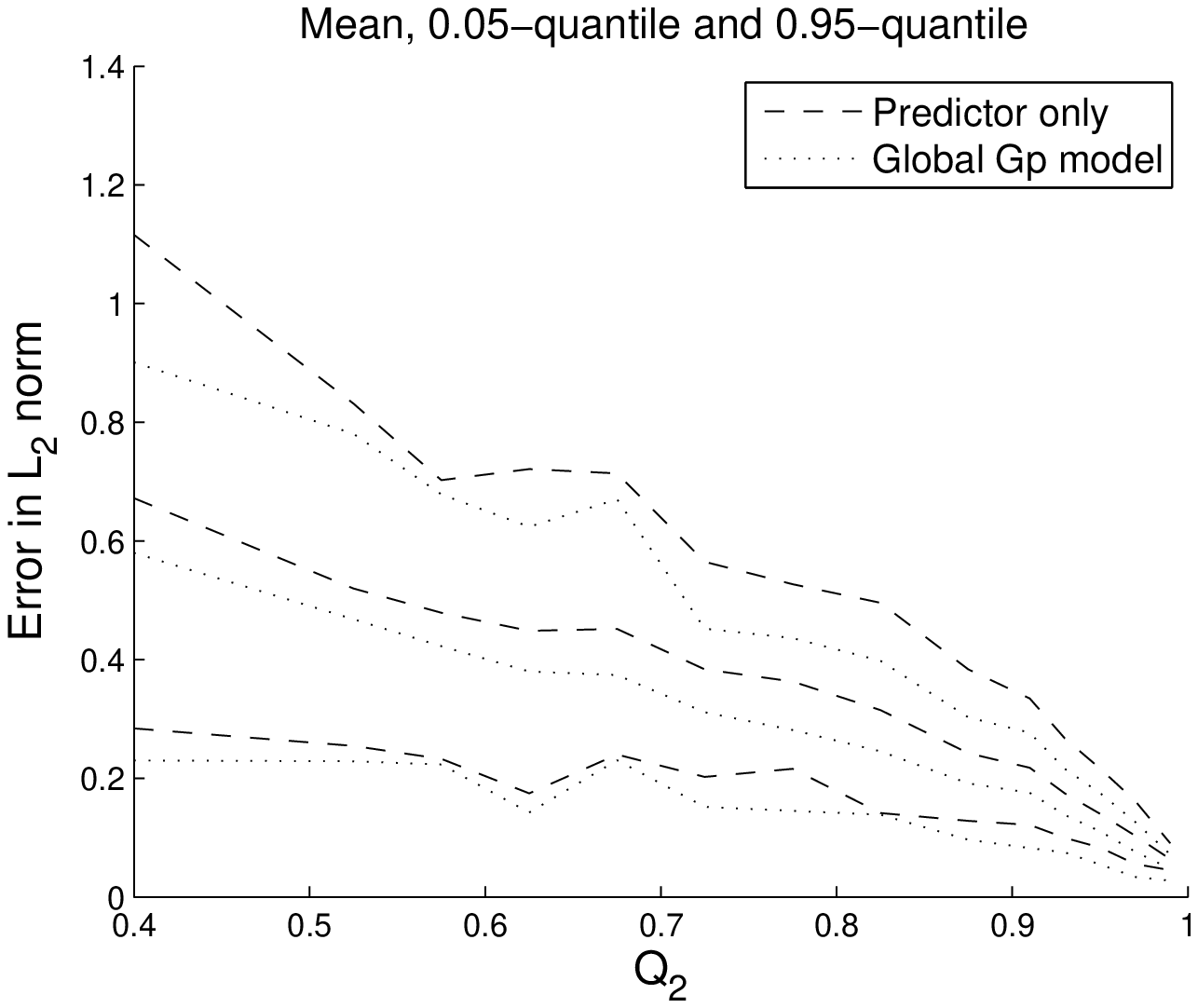} 
\end{tabular}
\caption{Error in $L_2$ norm for sensitivity indices in function of $n$ and $Q_2$ (g-Sobol function).}
\label{fig_gSobol_L2}
\end{figure}

To sum up the convergence of the indices for the different inputs, it can be useful to consider the error between the theoretical values of Sobol indices $S_i^{theo}$ and the estimated ones in $L_2$ norm:
\begin{equation}\label{L2norme}
\text{err}_{L_2} = \sum _{i=1}^{d}(S_i^{theo}-\hat{S}_i)^2
\end{equation}
where $\hat{S}_i$ denotes the indices estimated with one of the two methods ($\hat{S}_i = S_i$ or $\hat{S}_i = \mu_{\tilde{S_i}}$~). 
Figure \ref{fig_gSobol_L2} illustrates this convergence in function of the learning sample size $n$ and in function of the predictivity coefficient $Q_2$.
  
%
%

From Figure \ref{fig_gSobol_L2}, we conclude that the sensitivity indices defined using the global Gp model ($\mu_{\tilde{S_i}}$) are better in mean than the one estimated with the predictor only ($S_i$). This difference between the two approaches is especially significant for high values of Sobol indices like the indices related to the first input ($S_1$ and $\mu_{\tilde{S_1}}$). For lower indices, these two approaches give in mean the same results. 
Even if the two sensitivity indices seem to have the same rate of convergence in function of $n$ or $Q_2$, it is important to notice that the second approach is more robust. Indeed, $\mu_{\tilde{S_i}}$ has a lower sampling deviation and variability than $S_i$. Besides, this higher robustness is more significant when the accuracy of the metamodel is weak ($Q_2<0.8$). So, taking into account the covariance structure of the Gp model appears useful to reduce the variability of the estimation of the sensitivity index.  

\subsection{Test on Ishigami function}\label{sec_comp_Ishig}

We now consider another analytical function currently used in sensitivity studies (Saltelli et al. \cite{salcha00}), the Ishigami function, where each of the three input random variables $(X_1,X_2, X_3)$ follows a uniform probability distribution on $\left[ -\pi,+\pi \right]$:
        \[
        f_{\mbox{\tiny Ishig}} (X_1,X_2, X_3) = \sin(X_1) + 7\sin^2(X_2)+ 0.1 X^{4}_3 \sin(X_1)
\]
The theoretical values of first order Sobol indices are known: 
\[
\left\{ \begin{array}{l}
 \displaystyle S_1 = 0.3139 \\
 \displaystyle S_2 = 0.4424 \\
 \displaystyle S_3 = 0 
   \end{array}  \right. 
  \]

\begin{figure}[ht]
\begin{tabular}{cc}
\includegraphics[height=7.cm,width=7.cm]{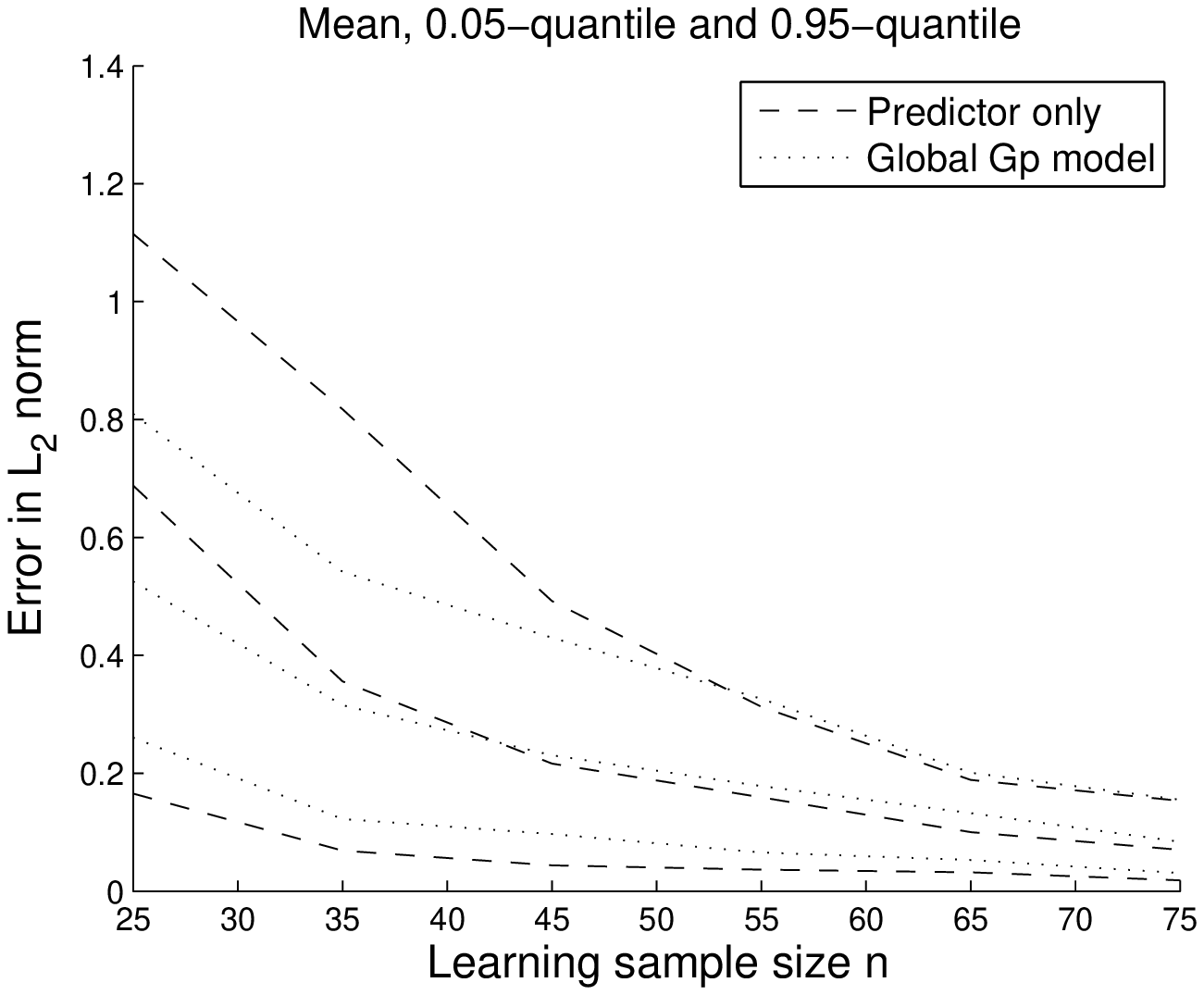} &
\includegraphics[height=7.cm,width=7.cm]{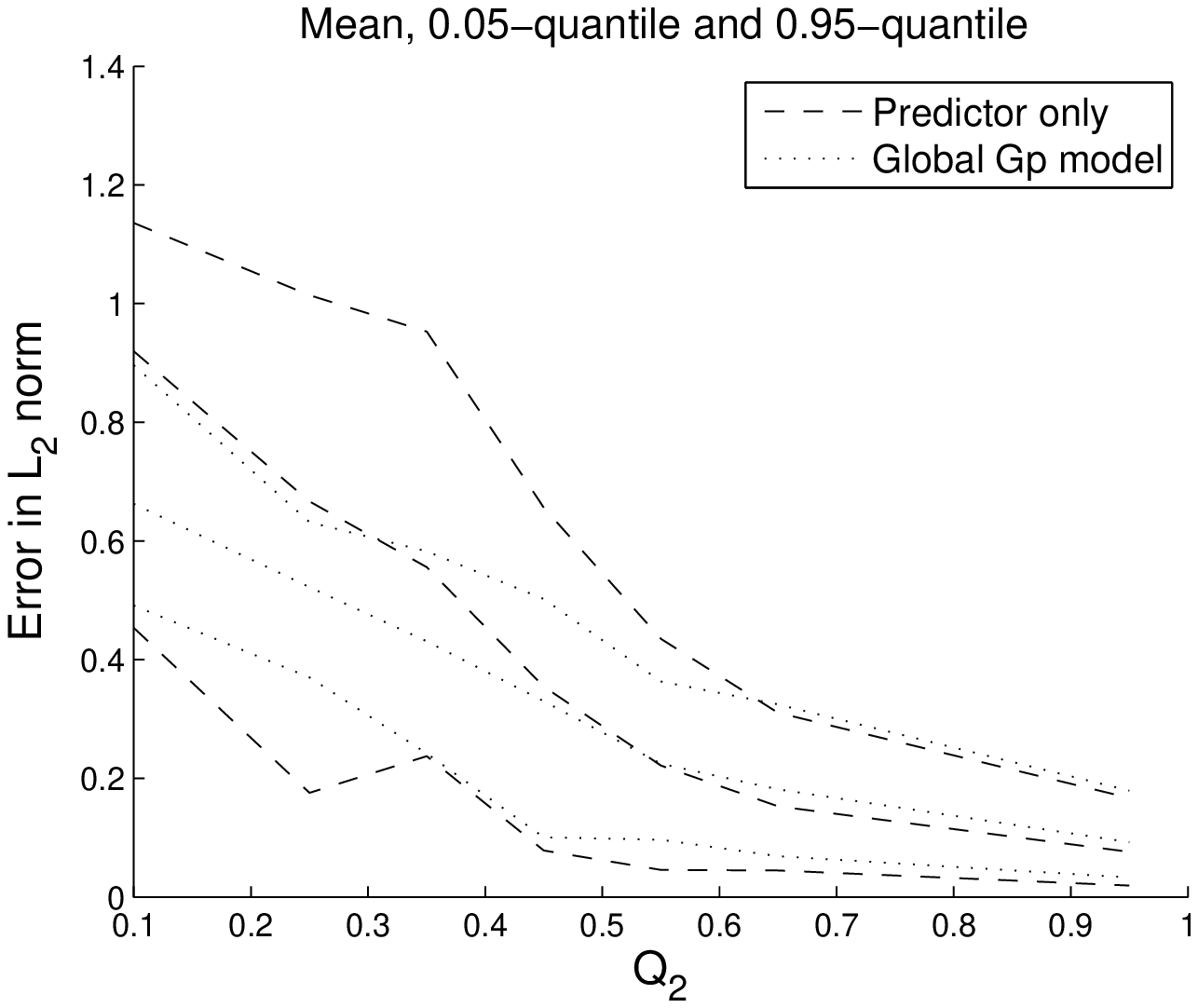} 
\end{tabular}
\caption{Error in $L_2$ norm for sensitivity indices in function of $n$ and $Q_2$ (Ishigami function).}
\label{fig_ishig_L2}
\end{figure}

Like for the g-function of Sobol, the error with the theoretical values of Sobol indices in $L_2$ norm is computed for the two approaches for different learning sample size $n$ and consequently for different values of $Q_2$. As before (Eq. (\ref{L2norme})), the diagrams of convergence are shown in figure \ref{fig_ishig_L2}.   

%

As observed for the g-function of Sobol, the indices defined with the global model are still more robust and less variable particularly for low values of $Q_2$. However, the difference between the mean of the two indices is not significant. For high values of the Gp model accuracy ($Q_2 > 0.8$), the two approaches give the same values but the first one (with only the predictor) remains easier to compute. 
So, the use of the covariance structure through the index $\tilde{S_i}$ seems to have a significant interest when the Gp metamodel is inaccurate or when few data are available to avoid too much variability of the estimated indices. 

\subsection{Construction of confidence intervals for sensitivity indices}

As well as being more robust in mean, the index defined with the second approach $\tilde{S_i}$ has the advantage to have a variance easy to compute. More generally, it is possible to build a confidence interval of any level for this sensitivity index, using the methodology described in section \ref{sec_sim_distrib} to simulate its distribution. This estimation of the uncertainty on the estimation of Sobol indices is another advantage of using the global Gp model: in practical cases with small learning sample size, only one Gp model is constructed.
The predictivity coefficient $Q_2$ can be estimated by cross-validation or leave-one-out, and if $Q_2$ shows a low predictivity (typically less than $0.8$), we wish to have some confidence in the estimation of Sobol indices computed from the Gp model.
Contrary to Gp model, other metamodels do not allow to directly estimate the Sobol indices uncertainties due to the model uncertainties.

To illustrate this, let us consider again the g-function of Sobol. Like in the previous section \ref{sec_comp_gSobol}, we consider different sizes of the learning sample (from $n=20$ to $n=50$). For each value of $n$, we build a conditional Gp model and we control its accuracy estimating the $Q_2$ on a test sample.
We simulate the distribution of $\tilde{S_i}$ to obtain the empirical $0.05$ and $0.95$-quantiles and consequently an empirical $90\%$-confidence interval.
Then, we check if the theoretical values of Sobol indices belong to the empirical $90\%$-confidence interval. We repeat this procedure $100$ times for each size $n$. 
Therefore, we are able to estimate the real level of our confidence interval and compare it to the $90\%$ expected. The real levels obtained in mean for any size $n$ and each input are presented in Table \ref{emp_level_gsobol}. 
\begin{table}[!ht]
\begin{center}
\begin{tabular}{cccc}
\hline
Variable & Theoretical value &  Mean of $\mu_{\tilde{S_i}}$  & Observed level of the empirical \\
   & of Sobol index &  & confidence interval \\
\hline
$X_1$ & 0.7164  & 0.7341 & 0.9381 \\
\hline
$X_2$ & 0.1791  & 0.1574 & 0.9369 \\
\hline
$X_3$ & 0.0237 & 0.0242 & 0.5830 \\
\hline
$X_4$ & 0.0072 & 0.0156 & 0.8886 \\
\hline
$X_5$ & 0.0001 & 0.0160 & 0.0674 \\
\hline
\end{tabular}
\end{center}
\caption{Real observed level of the empirical $90\%$-confidence interval built with the Gp model for the Sobol index of each input parameter (g-Sobol function).}
\label{emp_level_gsobol}
\end{table}  

For the high values of Sobol indices ($S_1$ and $S_2$ for example), the observed levels of the $90\%$-confidence interval built from the simulation of the distribution of  $\tilde{S_i}$ are really satisfactory and close to the expected level. In this case, the use of the global Gp model which gives confidence intervals for Sobol indices has a significant interest. On the other hand, for very low indices (close to zero), the Gp metamodel overestimates the Sobol indices.
It explains the inaccuracy of the confidence interval. Indeed, without a procedure of inputs selection, each variable appears in the Gp metamodel and in its covariance. Taking into account the variance leads to give a minimal bound for the influence of all the variables and consequently to overestimate the lowest Sobol indices. This tendency is confirmed by the observation of the mean of $\mu_{\tilde{S_i}}$ estimated for the three last inputs in Table \ref{emp_level_gsobol}. 
 
\begin{figure}[ht]
\begin{tabular}{ccc}
\includegraphics[height=5.cm,width=5.cm]{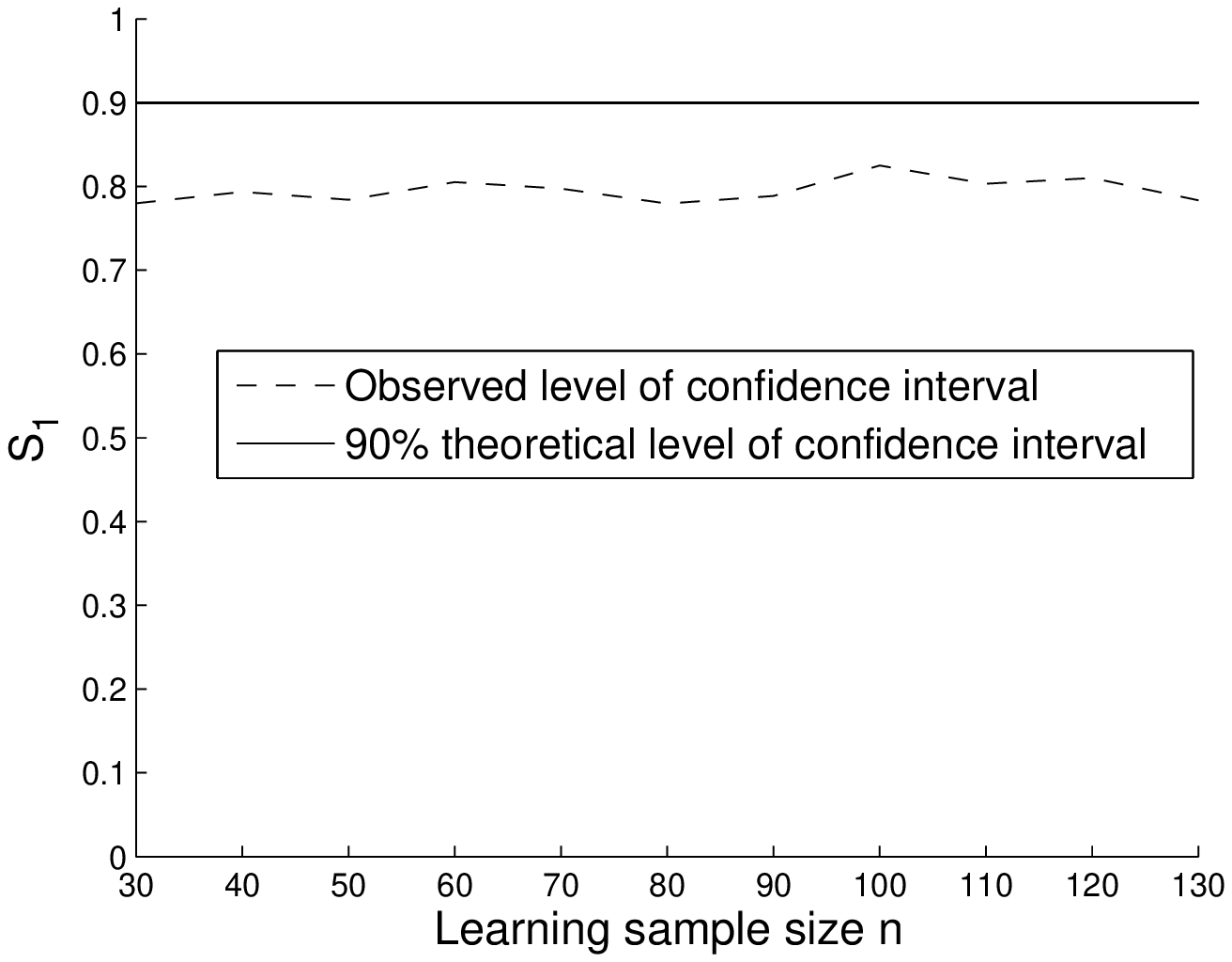} &
\includegraphics[height=5.cm,width=5.cm]{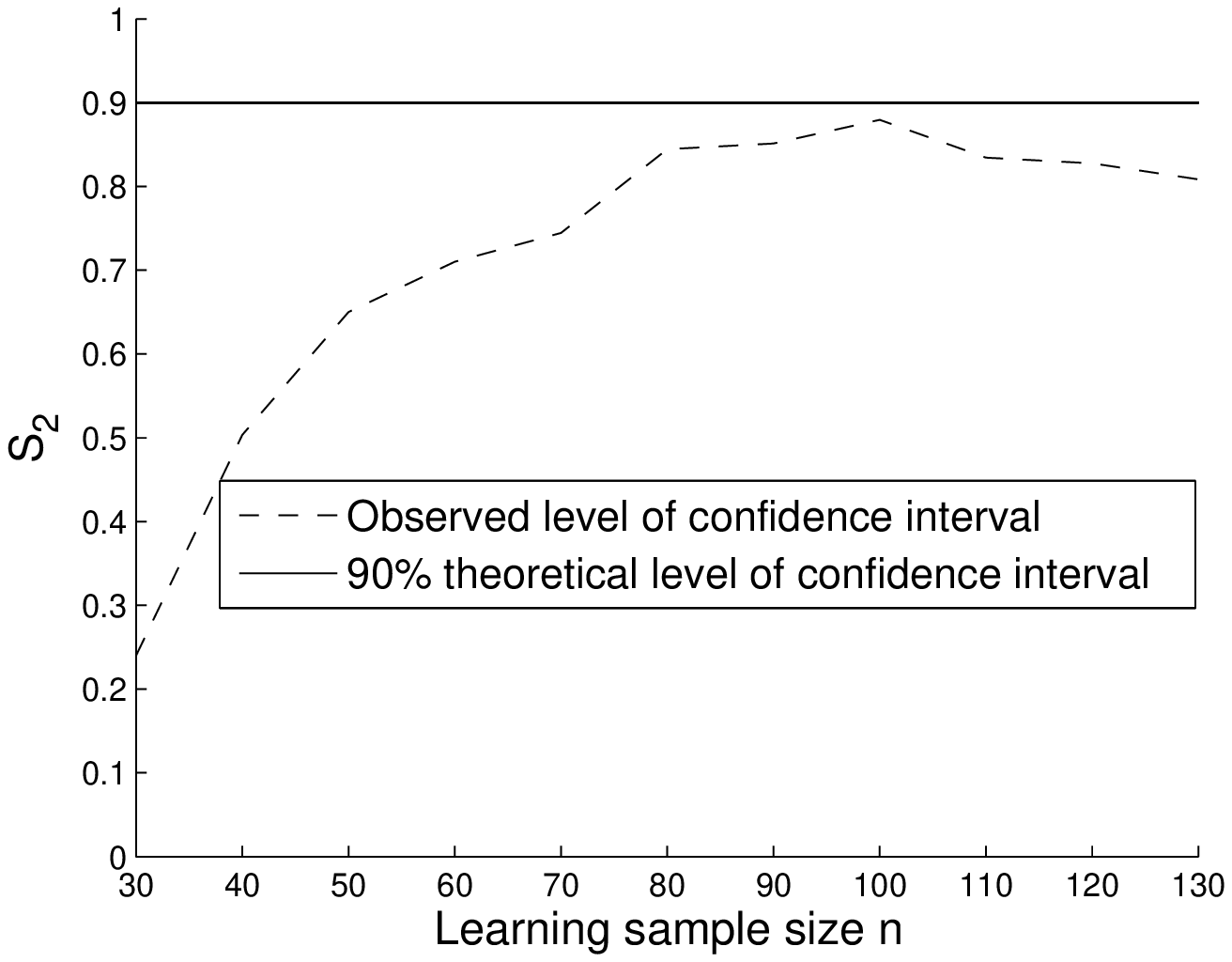} &
\includegraphics[height=5.cm,width=5.cm]{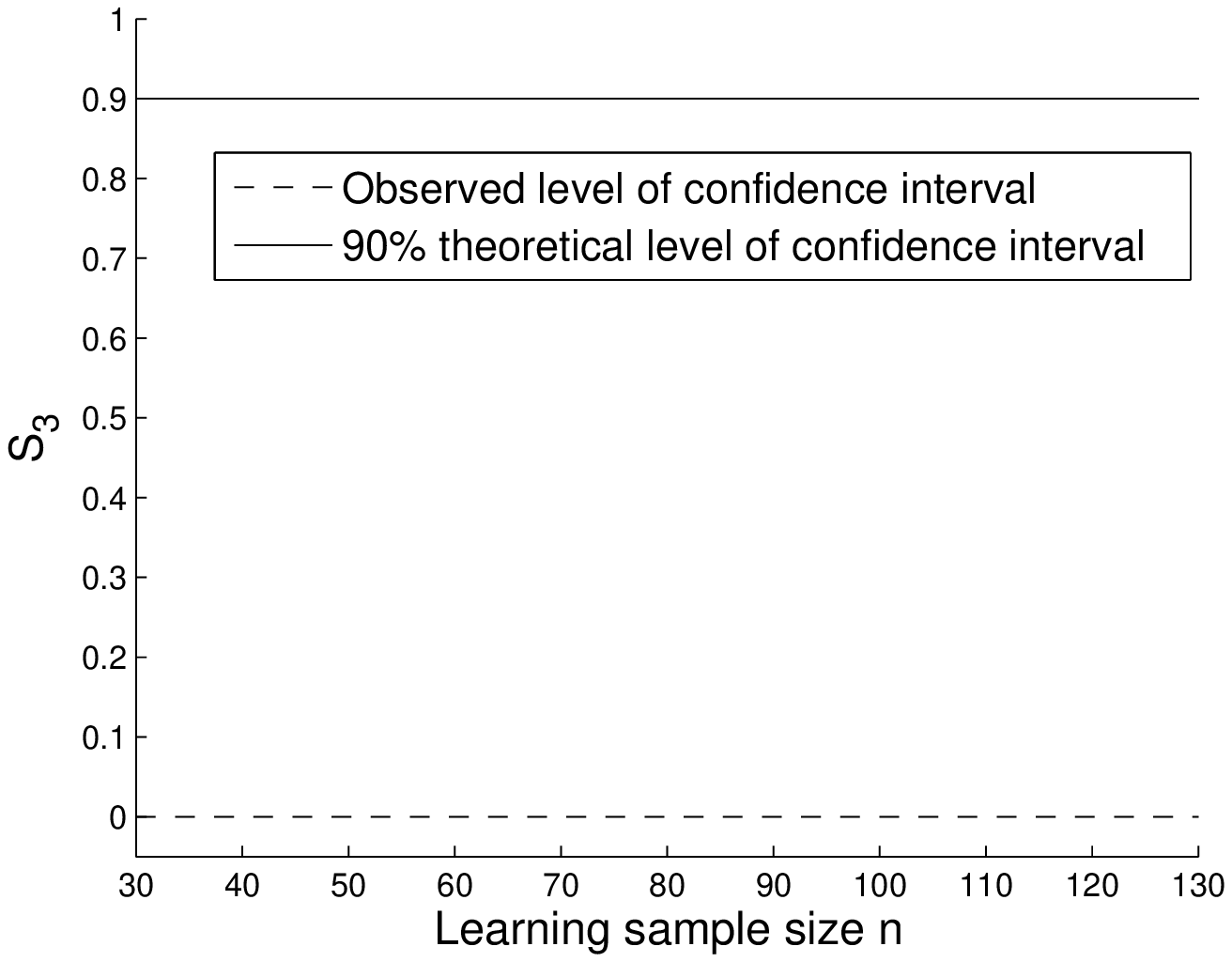} \\
\includegraphics[height=5.cm,width=5.cm]{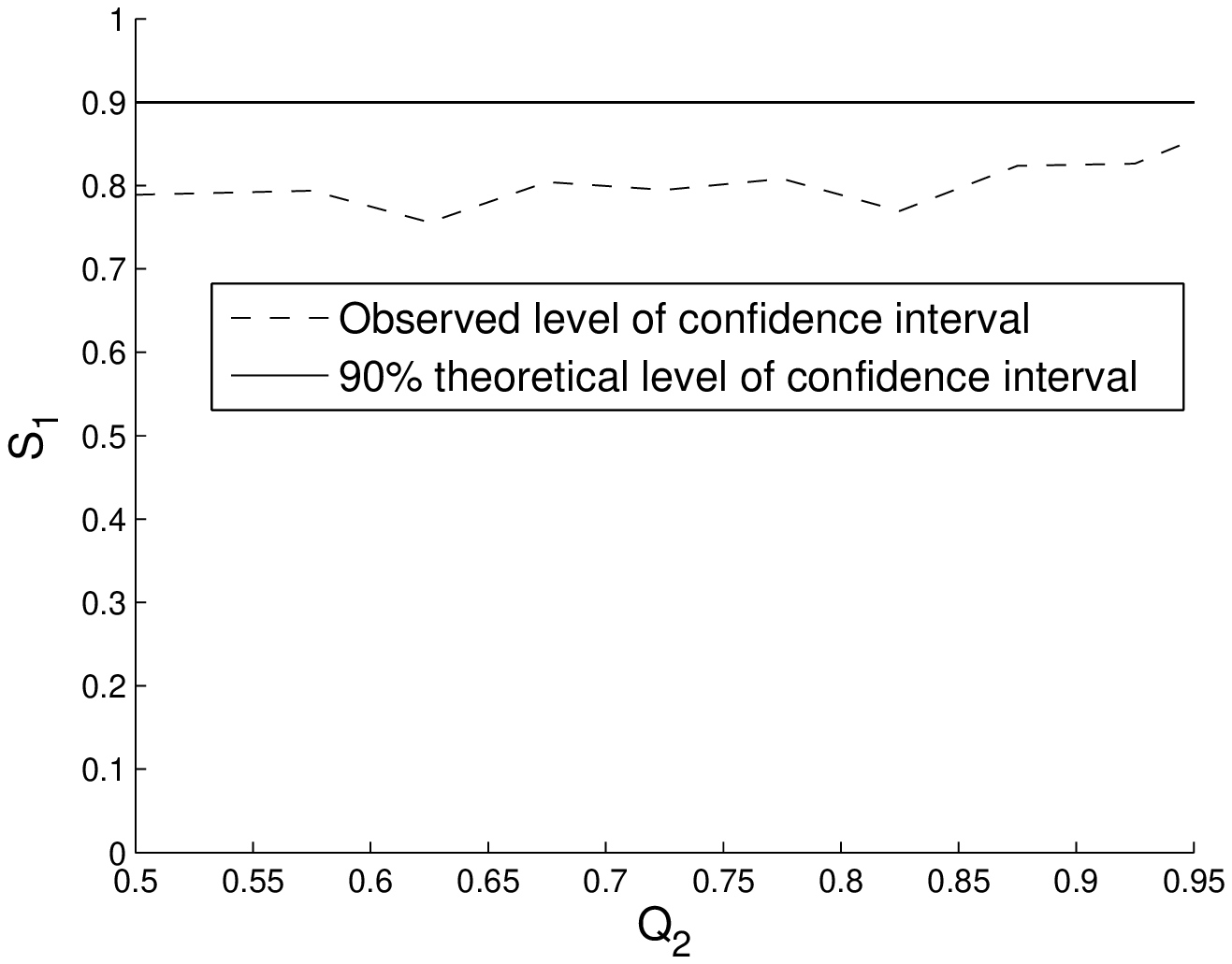} &
\includegraphics[height=5.cm,width=5.cm]{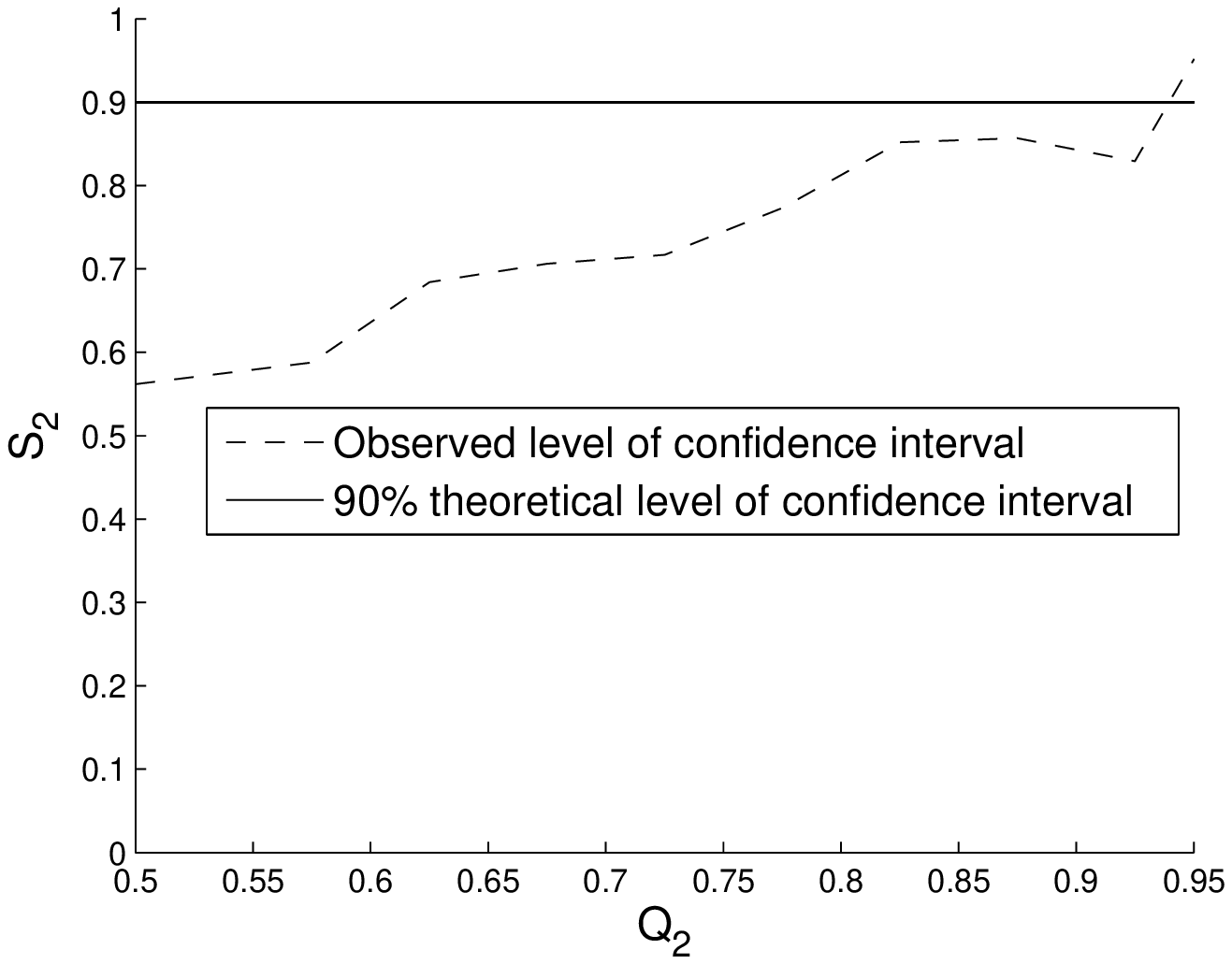} &
\includegraphics[height=5.cm,width=5.cm]{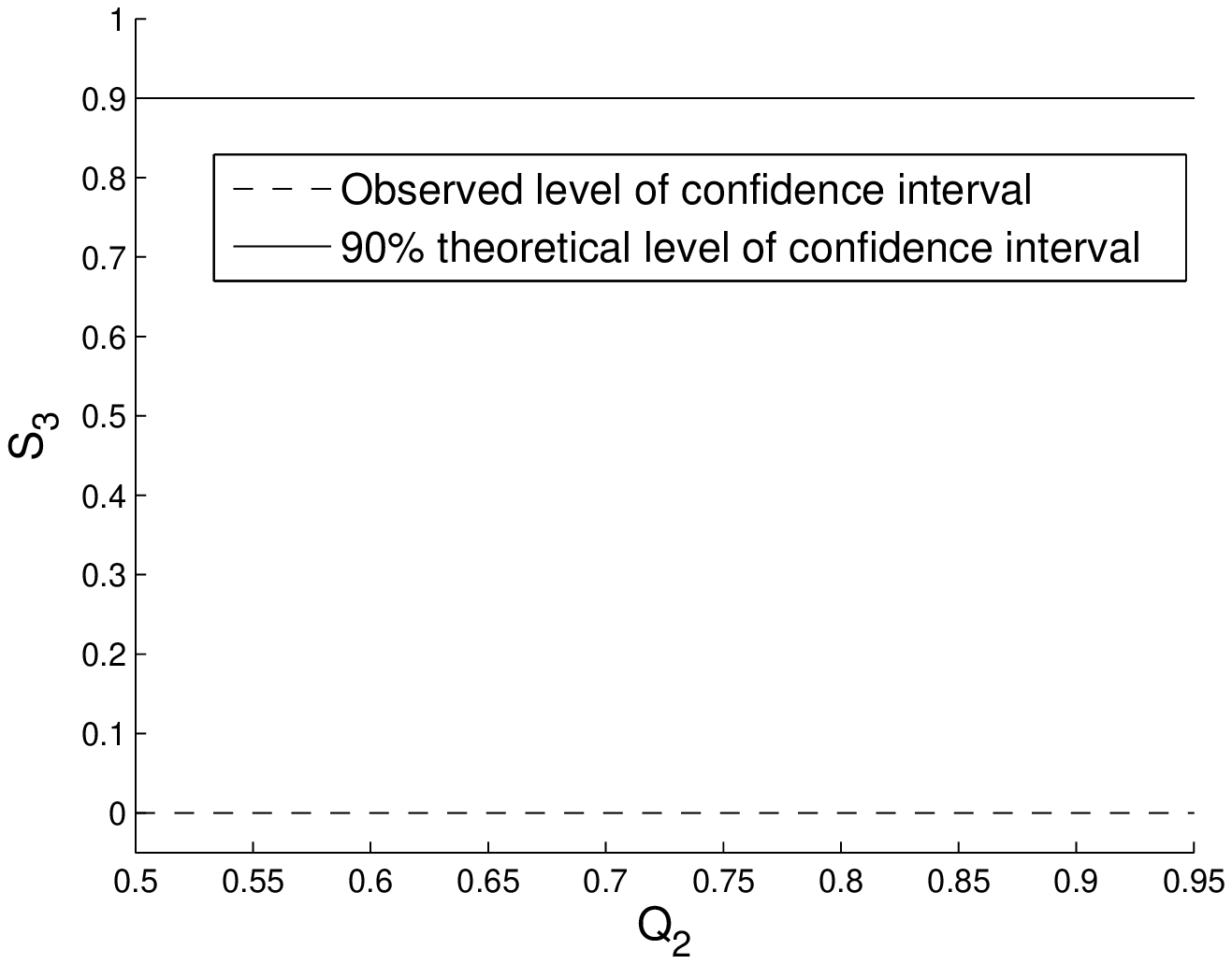} 
\end{tabular}
\caption{Convergence of the observed level of the empirical $90\%$-confidence in function of $n$ and $Q_2$ (Ishigami function).}
\label{fig_gsobol_IC}
\end{figure}

We can make the same study with the Ishigami function for $n = 30 $ to $n = 130$ which induces a $Q_2$ varying from $0.5$ to $0.95$. As all the procedure (i.e. learning sample simulation, Gp modeling and sensitivity analysis) is repeated $100$ times for each size $n$, the convergence of the observed level of the empirical $90\%$-confidence interval can be observed in function of $n$. Similarly, we can study this convergence in function of $Q_2$. Figure \ref{fig_gsobol_IC} shows all these diagrams of convergence. 
As previously remarked on the g-function of Sobol, the $90\%$-confidence intervals are efficient for the high values of Sobol indices ($S_1$ and $S_2$ for example). For these indices, the observed level of the confidence interval converges to theoretical level $0.9$. 
We can also notice that the predictivity quality of the Gp modeling which is required to obtain accurate confidence interval corresponds approximately to $Q_2 > 0.80$.
However, we judge that for $Q_2>0.6$, the error is not too strong and the obtained $90\%$-confidence interval
can be considered as a reliable and useful information.
On the other hand, for very low indices (close to zero), the problem of overestimating the Sobol indices still damages the accuracy of the interval confidence for any size $n$ and any $Q_2$. This remark is particularly true when the index is equal to zero (for example $S_3$).

\subsection{Application on an hydrogeologic transport code}

The  two approaches to compute the Sobol indices are now applied to the data obtained from the modeling of strontium 90 (noted $^{90}$Sr) transport in saturated porous media using the MARTHE software (developed by BRGM, France).
The MARTHE computer code models flow and transport equations in three-dimensional porous formations.
In the context of an environmental impact study, the MARTHE computer code has been applied to the model of $^{90}$Sr transport in saturated media for a radwaste temporary storage site in Russia (Volkova et al. \cite{volioo07}).
One of the final purposes is to determine the short-term evolution of $^{90}$Sr transport in soils in order to help the rehabilitation decision making. Only a partial characterization of the site has been made and, consequently, values of the model input parameters are not known precisely.
One of the first goals is to identify the most influential parameters of the computer code in order to improve the characterization of the site in a judicious way. 
To realize this global sensitivity analysis and because of large computing time of the MARTHE code, a Gp metamodel is built on the basis of a first learning sample.

\subsubsection{Data presentation} \label{subsecpres_appli}

The $20$ uncertain model parameters are permeability of different geological layers composing the simulated field (parameters 1 to 7), longitudinal dispersivity coefficients (parameters 8 to 10), transverse dispersivity coefficients (parameters 11 to 13), sorption coefficients (parameters 14 to 16), porosity (parameter 17) and meteoric water infiltration intensities (parameters 18 to 20). To study sensitivity of the MARTHE code to these parameters, $300$ simulations of these $20$ parameters have been made by the LHS method. 
For each simulated set of parameters, MARTHE code computes transport equations of $^{90}$Sr and predicts the evolution of $^{90}$Sr concentration for year 2010. For each simulation, the output we consider is the value of $^{90}$Sr concentration, predicted for year 2010, in a piezometer located on the waste repository site.

\subsubsection{Gp modeling and computation of Sobol indices}
To model the concentration in the piezometer predicted by MARTHE code in 2010 in function of the $20$ input parameters, we fit a Gp metamodel conditionally to $300$ simulations of the code. The regression and correlation parameters of the Gp model are estimated by maximum likelihood and a procedure of input selection is used. The input variables introduced in the metamodel are the sorption coefficient of the upper layer (parameter 14 denoted $kd1$), an infiltration intensity (parameter 20 denoted $i3$) and the permeability of the upper layer (parameter 1  denoted $per1$). The accuracy of the Gp model is checked with the estimation of $Q_2$ by a cross validation on the learning sample. The predictivity coefficient estimated is: $Q_2 = 0.92$. From previous study (Marrel et al. \cite{marioo07}), we have found that the linear regression gives a $Q_2 = 0.69$ and the metamodel based on boosting of regression trees gives a $Q_2 = 0.83$.
From laboratory measures and bibliographical information, prior distributions have been determined for the inputs $kd1$, $i3$ and $per1$ and are respectively a Weibull, a trapezoidal and a uniform distributions. The parameters of these distributions has been estimated or fixed a priori. 
Then, using the global Gp model, the Sobol indices defined by $\mu_{\tilde{S_i}}$ are computed (Eq. (\ref{eq_sobol_approach2b})) as well as the standard deviation $\sigma_{\tilde{S_i}}$ and the $90\%$-confidence interval associated (cf. methodology \ref{sec_sim_distrib}). The results are presented in Table \ref{sobol_piezo}, with the Sobol indices obtained with the predictor-only approach
and with the boosting predictor. 
   \begin{table}[ht]
\begin{center}
\begin{tabular}{cccccc}
\hline
input variable & Boosting of & Predictor only & \multicolumn{3}{c}{Whole Gp model} \\
& regression trees &  (Gp model)  & & & \\
\cline{4-6}
  & ${S_i}$ & ${S_i}$ & $\mu_{\tilde{S_i}}$ & $\sigma_{\tilde{S_i}}$ & $90\%$-confidence interval \\
\hline
per1 & 0.03 & 0.081 & 0.078 & 0.020  & $\left[ \; 0.046 \: ; \: 0.112 \; \right]$ \\
\hline
kd1 & 0.90 & 0.756 & 0.687  & 0.081  &  $\left[ \; 0.562 \: ; \: 0.825 \; \right]$ \\
\hline
i3 & 0.03 & 0.148 & 0.132 & 0.022 &  $\left[ \; 0.100 \: ; \: 0.170 \; \right]$ \\
\hline
\end{tabular}
\caption{Estimated Sobol indices, associated standard deviation and confidence intervals for MARTHE data.}\label{sobol_piezo}
\end{center}
\end{table}
The use of Gp model gives a better predictivity than the boosting of regression trees (respectively $Q_2 = 0.92$ and $Q_2 = 0.83$) and consequently a more accurate estimation of Sobol indices. Besides, the Sobol indices estimated with the boosting model do not even belong to the confidence intervals given by the Gp model. Even if the sensitivity indices based on the predictor only and the ones estimated with the whole Gp model are very close, the second approach has the advantage to give confidence intervals and consequently to have a more rigorous analysis. 

Without considering their interactions, the 3 inputs $kd1$, $i3$ and $per1$ explained nearly $90 \%$ of the total variance of the output and the most influent input is clearly $kd1$, followed by $i3$ and $per1$. So, $kd1$ is the most important parameter to be characterized in order to reduce the variability of the concentration predicted by MARTHE code. 
Using the whole Gp model, we also have an indication of the accuracy of Sobol indices. The standard deviation of the indices are small and increase the confidence in the value estimated ($\mu_{\tilde{S}_{kd1}}$, $\mu_{\tilde{S}_{i3}}$ and $\mu_{\tilde{S}_{per1}}$). Moreover, the very small overlap of the $90\%$-confidence interval of the 3 indices indicates that the order of influence of the inputs is well determined and strongly confirms the predominance of $kd1$.
So, the confidence intervals and the standard deviation obtained with the whole Gp model give more confidence in the interpretation of Sobol indices. 

Taking into account the variability of the Gp model via its covariance structure gives more robustness to the results and their analysis.
However, this increase of precision and confidence has a numerical cost. Indeed, the number of numerical integrals being computed is of order $O(dn^2)$ where $d$ is the number of inputs and $n$ the number of simulations, i.e. the learning sample size. The numerical cost depends also on the numerical precision required for the approximation of the integrals. Moreover, a high precision is often essential to provide the robustness of the computation of Sobol indices, especially when the distribution of the inputs is narrow and far from the uniform distribution (like the Weibull distribution of $kd1$). In this last case, it can be judicious to adapt the numerical scheme in order to increase the precision in the region of high density.   

\section{CONCLUSION} \label{secdisc}

We have studied the Gaussian process metamodel to perform sensitivity analysis, by estimating Sobol indices, of complex computer codes. 
 This metamodel is built conditionally to a learning sample, i.e. to $n$ simulations of the computer code.
 The Gp model proposes an analytical formula which can be directly used to derive analytical expressions of Sobol indices. Indeed, in the case of independent inputs and with our choice of regression and covariance functions, the formula of Gp model leads to one and two-dimensional numerical integrals, avoiding a large number of metamodel predictor evaluations in Monte Carlo methods. 
The use of Gp model instead of other metamodel is therefore highly efficient. 
 Another advantage of Gp metamodel stands in using its covariance structure to compute Sobol indices and to build associated confidence intervals, by using the global stochastic model including its covariance. 

On analytical functions, the behavior and convergence of the Sobol index estimates were studied in function of the learning sample size $n$ and the predictivity of the Gp metamodel. This analysis reveals the significant interest of the global stochastic model approach when the Gp metamodel is inaccurate or when few data are available. Indeed, the use of the covariance structure gives sensitivity indices which are more robust and less variable.
Moreover, all the distribution of the sensitivity index (defined as a random variable) can be simulated following an original algorithm. Confidence intervals of any level for the Sobol index can then be built. In our tests, the observed level of the interval was compared to the expected one on analytical functions. For the highest values of Sobol indices and under the hypothesis of a Gp metamodel with a predictivity coefficient larger than $60\%$, the confidence intervals are satisfactory. In this case, the use of the global Gp model which gives confidence intervals for Sobol indices has a significant interest. The only drawback is that the use of covariance structure has a tendency to give a minimal bound for the influence of all the variables and consequently to overestimate the lowest Sobol indices and to give inaccurate confidence intervals for very low indices (close to zero).

The use of covariance structure was also illustrated on real data, obtained from a complex hydrogeological computer code, simulating radionuclide groundwater transport. This application confirmed the interest of the second approach and the advantage of Gp metamodel which, unlike other efficient metamodels (neural networks, regression trees, polynomial chaos, \ldots), gives confidence intervals for the estimated sensitivity indices. 
The same approach based on the use of the global Gp metamodel can be used to make uncertainty propagation studies and to estimate the distribution of the computer code output in function of the uncertainties on the inputs.


\section{ACKNOWLEDGMENTS}     

This work was supported by the MRIMP project of the ``Risk Control Domain'' that is managed by CEA/Nuclear Energy Division/Nuclear Development and Innovation Division. We are also grateful to S\'ebastien da Veiga for helpful discussions.



\singlespacing
\bibliographystyle{plain}  
\bibliography{bibl_hal}

\end{document}